\documentclass[12pt,preprint]{aastex}

\usepackage{lineno}
\linenumbers

\shorttitle{{\it Suzaku} Observations of Luminous Quasars}
\shortauthors{Sato et al.}

\begin{document}

\title{{\it Suzaku} Observations of Luminous Quasars:\\ 
Revealing the Nature of High-Energy Blazar Emission\\ 
in Low-level activity States}

\author{
A.~A.~Abdo\altaffilmark{1,2}, 
M.~Ackermann\altaffilmark{3}, 
M.~Ajello\altaffilmark{3}, 
E.~Antolini\altaffilmark{4,5}, 
L.~Baldini\altaffilmark{6}, 
J.~Ballet\altaffilmark{7}, 
G.~Barbiellini\altaffilmark{8,9}, 
M.~G.~Baring\altaffilmark{10}, 
D.~Bastieri\altaffilmark{11,12}, 
K.~Bechtol\altaffilmark{3}
% ...
}
\altaffiltext{1}{Space Science Division, Naval Research Laboratory, Washington, DC 20375, USA}
\altaffiltext{2}{National Research Council Research Associate, National Academy of Sciences, Washington, DC 20001, USA}
\altaffiltext{3}{W. W. Hansen Experimental Physics Laboratory, Kavli Institute for Particle Astrophysics and Cosmology, Department of Physics and SLAC National Accelerator Laboratory, Stanford University, Stanford, CA 94305, USA}
\altaffiltext{4}{Istituto Nazionale di Fisica Nucleare, Sezione di Perugia, I-06123 Perugia, Italy}
\altaffiltext{5}{Dipartimento di Fisica, Universit\`a degli Studi di Perugia, I-06123 Perugia, Italy}
\altaffiltext{6}{Istituto Nazionale di Fisica Nucleare, Sezione di Pisa, I-56127 Pisa, Italy}
\altaffiltext{7}{Laboratoire AIM, CEA-IRFU/CNRS/Universit\'e Paris Diderot, Service d'Astrophysique, CEA Saclay, 91191 Gif sur Yvette, France}
\altaffiltext{8}{Istituto Nazionale di Fisica Nucleare, Sezione di Trieste, I-34127 Trieste, Italy}
\altaffiltext{9}{Dipartimento di Fisica, Universit\`a di Trieste, I-34127 Trieste, Italy}
\altaffiltext{10}{Rice University, Department of Physics and Astronomy, MS-108, P. O. Box 1892, Houston, TX 77251, USA}
\altaffiltext{11}{Istituto Nazionale di Fisica Nucleare, Sezione di Padova, I-35131 Padova, Italy}
\altaffiltext{12}{Dipartimento di Fisica ``G. Galilei'', Universit\`a di Padova, I-35131 Padova, Italy}
% ...

\begin{abstract}
We present the results from the {\it Suzaku} X-ray observations of five flat-spectrum radio 
quasars (FSRQs), namely PKS\,0208$-$512, Q\,0827+243, PKS\,1127$-$145, PKS\,1510$-$089 and 
3C\,454.3. All these sources were additionally monitored simultaneously or quasi-simultaneously 
by the {\it Fermi} satellite in gamma-rays and the {\it Swift} UVOT in the UV and optical 
bands, respectively. We constructed their broad-band spectra 
covering the frequency range from $10^{14}$\,Hz up to $10^{25}$\,Hz, 
and those reveal the nature of high-energy emission of luminous blazars in their low-activity states. The analyzed X-ray spectra are well fitted by a power-law model with 
photoelectric absorption. In the case of PKS\,0208$-$512, PKS\,1127$-$145, and 3C\,454.3, 
the X-ray continuum showed indication of hardening at low-energies.
Moreover, when compared with the previous X-ray observations, we see a significantly increasing
contribution of low-energy photons to the total X-ray fluxes when the sources are getting
fainter. The same behavior can be noted in the {\it Suzaku} data alone. A likely explanation 
involves a variable, flat-spectrum component produced via inverse-Compton (IC) emission,
plus an additional, possibly steady soft X-ray component prominent when the source gets 
fainter. This soft X-ray excess is represented either by a steep power-law (photon indices
$\Gamma \sim 3-5$) or a blackbody-type emission with temperatures $kT\sim0.1-0.2$\,keV. 
We model the broad-band spectra spectra of the five observed FSRQs using synchrotron self-Compton 
(SSC) and/or external-Compton radiation (ECR) models. Our modeling suggests that the
difference between the low- and high-activity states in luminous blazars is due to
the different total kinetic power of the jet, most likely related to varying bulk Lorentz factor 
of the outflow within the blazar emission zone.
\end{abstract}


\section{Introduction}


Observations with the EGRET instrument ($30$\,MeV to $30$\,GeV; Thompson et al. 1993)
on board the Compton Gamma-Ray Observatory (CGRO) have resulted in detection of 
$\gamma$-ray emission from a few hundred astrophysical sources, 66 of which
were securely associated with active galactic nuclei (AGNs; e.g., Hartman et al. 1999).
Most of the AGNs detected by EGRET show characteristics of the blazar class. 
Observationally, this class include flat-spectrum radio quasars (FSRQs) 
and BL Lac objects. FSRQs have strong and broad optical emission lines, while the 
lines are weak or absent in BL Lacs. During the first three 
months of the {\it Fermi} Large Area Telescope's (LAT) all-sky-survey, 132 bright 
sources at high Galactic latitudes ($|b|>10^{\circ}$) were detected at a confidence level 
greater than $10\,\sigma$ (Abdo et al. 2009a). As expected from the EGRET observations, 
a large fraction (106) of these sources have been associated with known AGNs 
(Abdo et al. 2009b). This includes two radio galaxies (Centaurus\,A and NGC\,1275; 
Abdo et al. 2009c) and 104 blazars consisting of 58 FSRQs, 42 BL Lac objects, and 
4 blazars with unknown classification based on their Spectral Energy Distribution (SED).  

The radio-to-optical emission of luminous blazars of the FSRQ type is known 
to be produced by the synchrotron radiation of relativistic electrons accelerated 
within the outflow, while the inverse Compton (IC) scattering of low-energy photons 
by the same relativistic electrons is most likely responsible for the 
formation of the high energy X-ray-to-$\gamma$-ray component.
In addition, it is widely believed that the IC emission from FSRQs is dominated 
by the scattering of soft photons external to the jet (external Compton radiation, ECR). 
These photons, in turn, are produced by the accretion disk, and interact with the jet 
either directly or indirectly, after being scattered or reprocessed in the broad-line 
region (BLR) or a dusty torus (DT; see, e.g., Dermer \& Schlickeiser 1993; Sikora et
al. 1994; B\l a\.zejowski et al. 2000). Other sources of seed photons can also contribute 
to the observed IC radiation, and these are in particular jet synchrotron photons
through the synchrotron self-Compton process (hereafter SSC; Maraschi et al. 1992; 
Sokolov \& Marscher 2005).

In this context, detailed X-ray studies offer a unique possibility for discriminating
between different proposed jet emission models, since those scenarios predict 
distinct components to be prominent in blazar spectra around keV photon energies.
For example, in the soft X-ray range a break is expected in the ECR/BLR model, 
tracking the low-energy end of the electron energy distribution (Tavecchio et al. 2000; Sikora et al. 2009). Indeed, both the {\it XMM-Newton} and the 
{\it Suzaku} X-ray data of RBS 315 show ``convex" spectra (Tavecchio et al. 2007).
Such a curvature, on the other hand, can be alternatively accounted for by an excess 
absorption below 1 keV over the Galactic value, or by an intrinsic curvature in the
electron energy distribution. Furthermore, the situation can be more complex, with 
the simultaneous presence of yet additional components, such as the high-energy tail 
of the synchrotron continuum, SSC emission, or the narrow-band spectral feature originating 
from the ``bulk Comptonization" of external UV (disk) radiation by cold electrons within 
the innermost parts of relativistic outflow (Begelman \& Sikora 1987; Sikora \& Madejski 
2000; Moderski et al. 2004; Celotti et al. 2007).


Ghisellini et al. (1998) have studied the spectral energy distribution of 
51 EGRET-detected $\gamma$-ray loud blazars and have applied the SSC+ECR 
model to the spectra of these sources. Although most of the broadband data 
collected by Ghisellini et al (1998) corresponded to non-simultaneous measurements, 
those authors discovered clear trends and correlations among the physical quantities 
obtained from the model calculations. In particular, they found an evidence for a 
well-defined sequence such that the observed spectral properties of different 
blazar classes (BL Lacs and FSRQs) can be explained by an increasing 
contribution of an external radiation field towards cooling jet electrons 
(thus producing the high-energy emission) with the increasing jet power. 
As a result, while the SSC process alone may account for the entire high-energy 
emission of low-power sources (BL Lacs), a significant contribution from the
ECR is needed to explain the observed spectra of high-power blazars (FSRQs).
Meanwhile, when focusing on one particular object, Mukherjee et al. 
(1999) reported that they found a similar trend in the different spectral states 
of PKS\,0528+134. They studied the sequence of flaring and low-flux states 
of the source and found that the SSC mechanism plays a more important role 
when the source is in a low state, and the ECR mechanism is the dominant 
electron cooling mechanism when the source is in a high $\gamma$-ray state
(see in this context also Sambruna et al. 1997).

In order to understand the blazar phenomenon and the differences between 
BL Lacs and FSRQs, as well as the origin of spectral transitions in a particular 
object, one has to obtain truly simultaneous coverage across the entire spectrum, 
during both flaring and low-activity states. However, past $\gamma$-ray 
observations in low-activity states have been limited to only a few extremely 
luminous objects, such as PKS\,0528-134 or 3C\,279. Only now,
with the successful launch of the {\it Fermi} satellite and the excellent performance
of the {\it Suzaku} instruments, do we have an opportunity to study high-energy spectra 
of blazars with substantially improved sensitivity, and therefore 
can probe the different states of the sources' activity.

In this paper, we report the high-sensitivity, broadband {\it Suzaku} observations of five FSRQs, 
namely PKS\,0208$-$512, Q\,0827+243, PKS\,1127$-$145, PKS\,1510$-$089, and 3C\,454.3, 
which were bright gamma-ray sources detected by EGRET. 
Additionally, all of these sources were monitored simultaneously or 
quasi-simultaneously by the {\it Fermi} LAT and {\it Swift} Ultraviolet/Optical Telescope 
(UVOT; Roming et al. 2005). These broadband and high-sensitivity observations allow us to reveal the 
characteristics of the high-energy IC continuum in the low-activity states 
of luminous blazars. The paper is organized as follows:  in $\S$2, we describe 
observation and data reduction in the X-ray ({\it Suzaku}), UV-optical ({\it Swift} 
UVOT) and $\gamma$-ray ({\it Fermi} LAT) domains. In $\S$3, we present the 
broad-band analysis results. Finally, in $\S$\,4 we discuss the constraints 
on the jet parameters and speculate on the the origin of different activity
states in luminous blazars. Throughout the paper we adopt the cosmological parameters 
$H_0 = 71$\,km\,s$^{-1}$\,Mpc$^{-1}$, $\Omega_{\rm M} = 0.27$, and $\Omega_{\Lambda} = 0.73$.

\section{Observation and Data Reduction}

\subsection{{\it Suzaku}}

Five FSRQs were observed by {\it Suzaku} (Mitsuda et al. 2007) for $40$\,ks each 
as one of the long-category projects between 2008 October and 2009 January. 
Table 1 summarizes the start time, end time, and the exposures for each observation. 
{\it Suzaku} carries four sets of X-ray telescopes (Serlemitsos et al. 2007),
each with a focal-plane X-ray CCD camera (XIS, X-ray Imaging Spectrometer;
Koyama et al. 2007) that is sensitive over the $0.3-12$\,keV band,
together with a non-imaging Hard X-ray Detector (HXD; Takahashi et al. 2007;
Kokubun et al. 2007), which covers the $10-600$\,keV energy band by utilizing Si PIN
photo-diodes and GSO scintillation detectors. All of the sources were focused 
on the nominal center position of the XIS detectors.

For the XIS, we used data sets processed using the software of the $Suzaku$ 
data processing pipeline (ver. 2.2.11.22). 
Reduction and analysis of the data were performed following the 
standard procedure using the HEADAS v6.5 software package.
The screening was based on the following criteria: 
(1) only ASCA-grade 0,2,3,4,6 events were accumulated, while hot 
and flickering pixels were removed using the CLEANSIS script, 
(2) the time interval after the passage of South Atlantic Anomaly 
was greater than $500$\,s, and (3) the object was at least $5^\circ$ and $20^\circ$ 
above the rim of the Earth (ELV) during night and day, respectively. 
In addition, we also selected the data with a cutoff rigidity (COR) larger 
than $6$\,GV. The XIS events were extracted from a circular region with a radius 
of $4.2^\prime$ centered on the source peak, whereas the background 
was accumulated in an annulus with inner and outer radii of 
$5.4^\prime$ and $7.3^\prime$, respectively. We checked that the use 
of different source and background regions did not affect the analysis 
results. The response and auxiliary files were produced using 
the analysis tools \textsc{xisrmfgen} and \textsc{xissimarfgen} 
developed by the {\it Suzaku} team, which are included in the software 
package HEAsoft version 6.5. 

The HXD/PIN data (version 2.0) were processed with basically the same screening
criteria as those for the XIS, except that we required ELV\,$\ge 5^\circ$ through
night and day and COR\,$\ge 8$\,GV. The HXD/PIN instrumental background 
spectra were provided by the HXD team for each observation (Kokubun et al. 
2007; Fukazawa et al. 2006). Both the source and background spectra were 
made with identical good time intervals and the exposure was corrected for 
detector deadtime of $6.0-8.0\%$. We used the response files, version 
\textsc{ae\_hxd\_pinhxdnom5\_20080716.rsp}, provided by the HXD team. 
In our analysis, the hard X-ray emission of PKS 1510$-$089 and 3C454.3 were 
detected in the energy range from 12 keV to 40 keV and 50 keV, respectively. 
For other objects, the sources were not detected in the HXD/PIN data.
We also note here that all of the objects, the sources were not detected in 
the HXD/GSO data.

\begin{table}[ht]
\footnotesize
\caption{{\it Suzaku} observation log of five FSRQs.}
\begin{center}
\label{table:pulsefit}
\begin{tabular}{ccccc}
\tableline
Object & $z$ & Start time & Stop time & XIS/HXD exposures \\
       &     & (UT)       & (UT)      & (ks)  \\
\tableline
0208$-$512 & 1.003 & 2008 Dec 14 07:33 & 2008 Dec 15 11:30 & 50.3/39.3 \\
0827+243 & 0.939 & 2008 Oct 27 05:11 & 2008 Oct 28 08:04 & 35.3/36.3 \\ 
1127$-$145 & 1.187 & 2008 Nov 29 18:10 & 2008 Nov 30 22:51 & 42.2/29.0 \\
1510$-$089 & 0.361 & 2009 Jan 27 04:32 & 2009 Jan 28 05:25 & 38.5/36.2 \\
3C\,454.3  & 0.859 & 2008 Nov 22 09:19 & 2008 Nov 23 16:31 & 39.9/40.4 \\
\tableline
\end{tabular}
\end{center}
\end{table}

\subsection{{\it Swift}}

Four analyzed FSRQs (PKS\,0208$-$512, Q\,0827+243, PKS\,1510$-$089, and 3C\,454.3) 
were observed with {\it Swift} between 2008 October and 2009 January, as part 
of {\it Swift} ``target of opportunity" observations. We analyzed the 
data taken within or near the time of the {\it Suzaku} 
observations. For the case of PKS\,1127$-$145, however, the observations were made only 
once in 2007 March. We focused on analysis of 
the UVOT data, since {\it Suzaku} provides much better 
photon statistics in X-rays than {\it Swift} X-ray Telescope (XRT; Burrows et al. 2005) 
and Burst Alert Telescope (BAT; Barthelmy et al. 2005), thanks to the long 
{\it Suzaku} exposures.  We used the 
XRT data primarily for a consistency check regarding the spectral properties.
Table \ref{table:Swiftobs} summarizes the start time, exposure time, and filters 
used for each observation. 

The UVOT observing mode commonly takes an exposure in each of the six optical and 
ultraviolet filters ($v$, $b$, $u$, uvw1, uvm2, and uvw2) per {\it Swift} pointing. 
The list of UVOT observations is given in Table \ref{table:Swiftobs}. 
For the screening, reduction and analysis of the {\it Swift} data, 
we used standard procedures within the HEASoft v.6.5 software package 
with the calibration database updated as of 2009 February 28. 
For this analysis, Level 2 sky-corrected image data were used. 
Since all sources were relatively bright, the source aperture sizes were 
chosen to correspond to those used to determine the UVOT zero points:
$5''$ for the optical and UV filters (Poole et al. 2008).
The background was extracted from a nearby source-free circular region with 
$15''$ radius. All image data were corrected for coincidence loss. 
The observed magnitudes were converted into flux densities by the 
standard procedures (Poole et al. 2008).

The XRT data were all taken in Photon Counting mode (PC mode; Hill et al. 2004).
The data were reduced by the XRT data analysis task \textsc{xrtpipline} version 0.12.0.
Photons were selected from the event file by xselect version 2.4.
The auxiliary response file was created by the XRT task \textsc{xrtmkarf} and the 
standard response file \textsc{swxpc0to12s6\_20010101v011.rmf}. 
All spectra were analyzed in the 0.3$-$10.0 keV band using XSPEC version 11.3.2.

\begin{table}[ht]
\footnotesize
\caption{{\it Swift} observation log of five FSRQs.}
\begin{center}
\label{table:Swiftobs}
\begin{tabular}{cccccc}
\tableline
Object & obsID & Start time & Exposure$^{a}$ & Exposure$^{b}$ & Filter$^{b}$ \\
       &       & (UT)       &   (ks)       &      (ks)       & \\
\tableline
0208$-$512 & 00035002024 & 2008 Dec 14 15:25 & 0.99   & 0.94 & all  \\
0827+243 & 00036375004 & 2008 Dec 08 13:42 & 1.72   & 1.71 & $u$ \\
1127$-$145 & 00036380001 & 2007 Mar 24 00:32 & 14.6   & 14.2 & all \\
1510$-$089 & 00031173010 & 2009 Jan 25 18:40 & 3.46   & 3.40 & $u$, w1, m2\\
3C\,454.3  & 00035030030 & 2008 Oct 26 20:28 & 0.43 & 0.40 & all\\
\tableline
\end{tabular}
\tablenotetext{a}{$Swift$ XRT}
\tablenotetext{b}{$Swift$ UVOT}
\end{center}
\end{table}

\subsection{{\it Fermi} LAT}

During the first year of {\it Fermi} Large Area Telescope (LAT; Atwood et al. 2009) 
operation, most of the telescope's time has been dedicated to ``survey mode'' 
observing, where {\it Fermi} points away from the Earth, and nominally rocks the 
spacecraft axis north and south from the orbital plane to enable monitoring of the 
entire sky every $\sim 3$ hours (or 2 orbits). We analyzed the LAT's observations 
of the five blazar regions using data collected during the first 4-5 months 
centered around {\it Suzaku} observations. Little variability indicated
by the LAT lightcurves for the studied objects during this time implies
that the constructed broad-band spectra, even though not exactly simultaneous,
are representative for the low-activity states of all five blazars.

The data used here comprise all scientific data obtained between 4 August 
and 19 December 2008 for PKS\,0208$-$512, Q\,0827+243, PKS\,1127$-$145 and 3C\,454.3 
(interval runs from Mission Elapsed Time (MET) 239557417 to 251345942), 
and 4 August 2008 and 30 January 2009 for 1510$-$089 (MET 239557417 to 254966035), 
respectively .
We have applied the zenith angle cut to eliminate photons from the Earth's limb, 
at $105^{\circ}$. This is important in pointed mode observations, but also 
important for survey mode due to overshoots and sun avoidance maneuvers. 
In addition, we excluded the time intervals when the rocking angle was 
more than 43$^{\circ}$.
We use the ``Diffuse" class events 
(Atwood et al. 2009), which, of all reconstructed events have the highest 
probability of being photons. 

In the analysis presented here, we set the lower energy bound to a value of 
$200$\,MeV, since the bin counts for photons with energies of $\sim 100$\,MeV 
and lower are systematically lower than expected based on extrapolations of a 
reasonable functions. Science Tools version v9r14 and IRFs (Instrumental 
Response Functions) P6\_V3 were used. 

\section{Results}

\subsection{{\it Suzaku}}

\subsubsection{Temporal analysis}

\begin{figure}[ht]
\begin{center}
\includegraphics[angle=0,scale=.3]{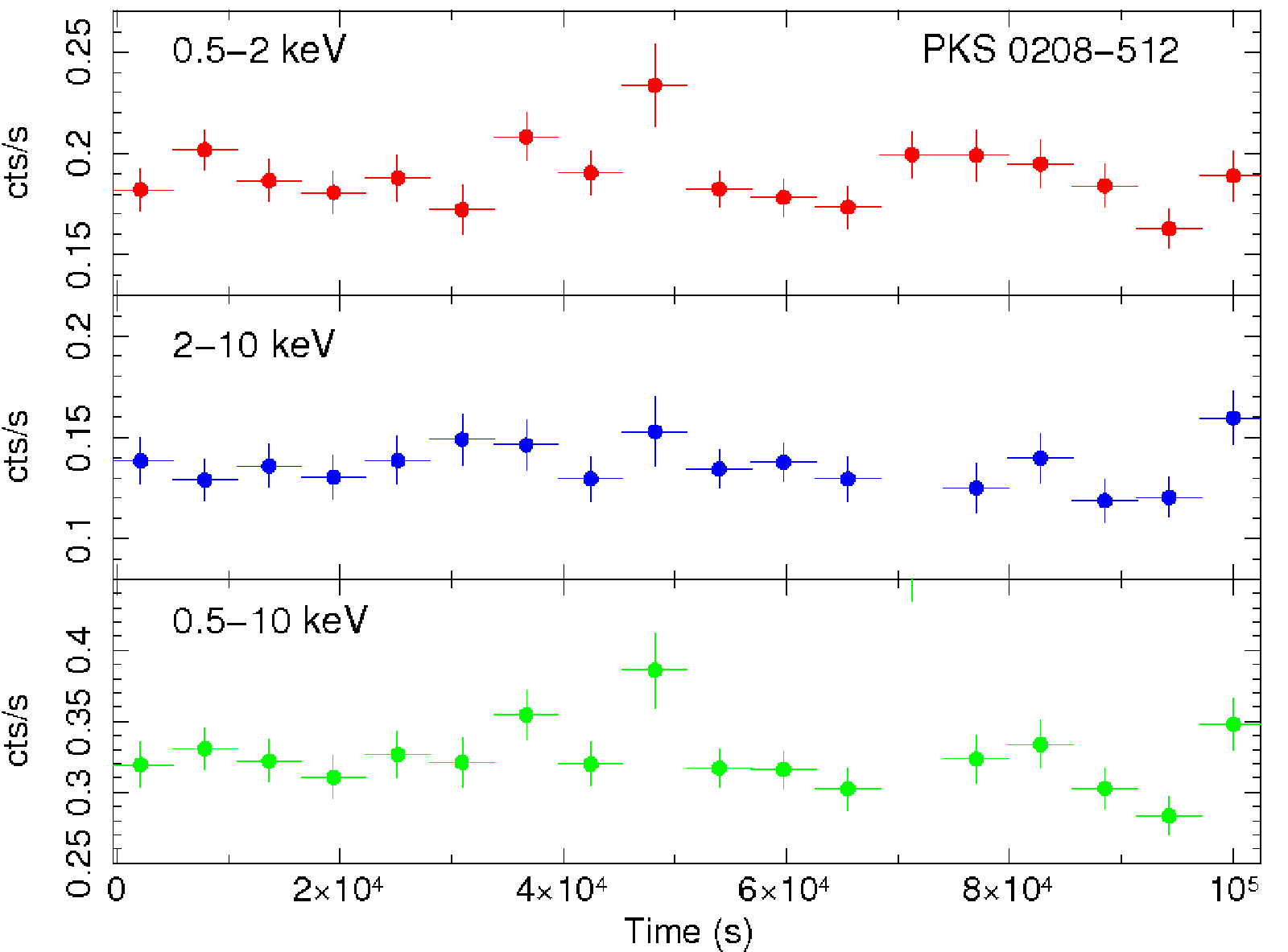}
\includegraphics[angle=0,scale=.3]{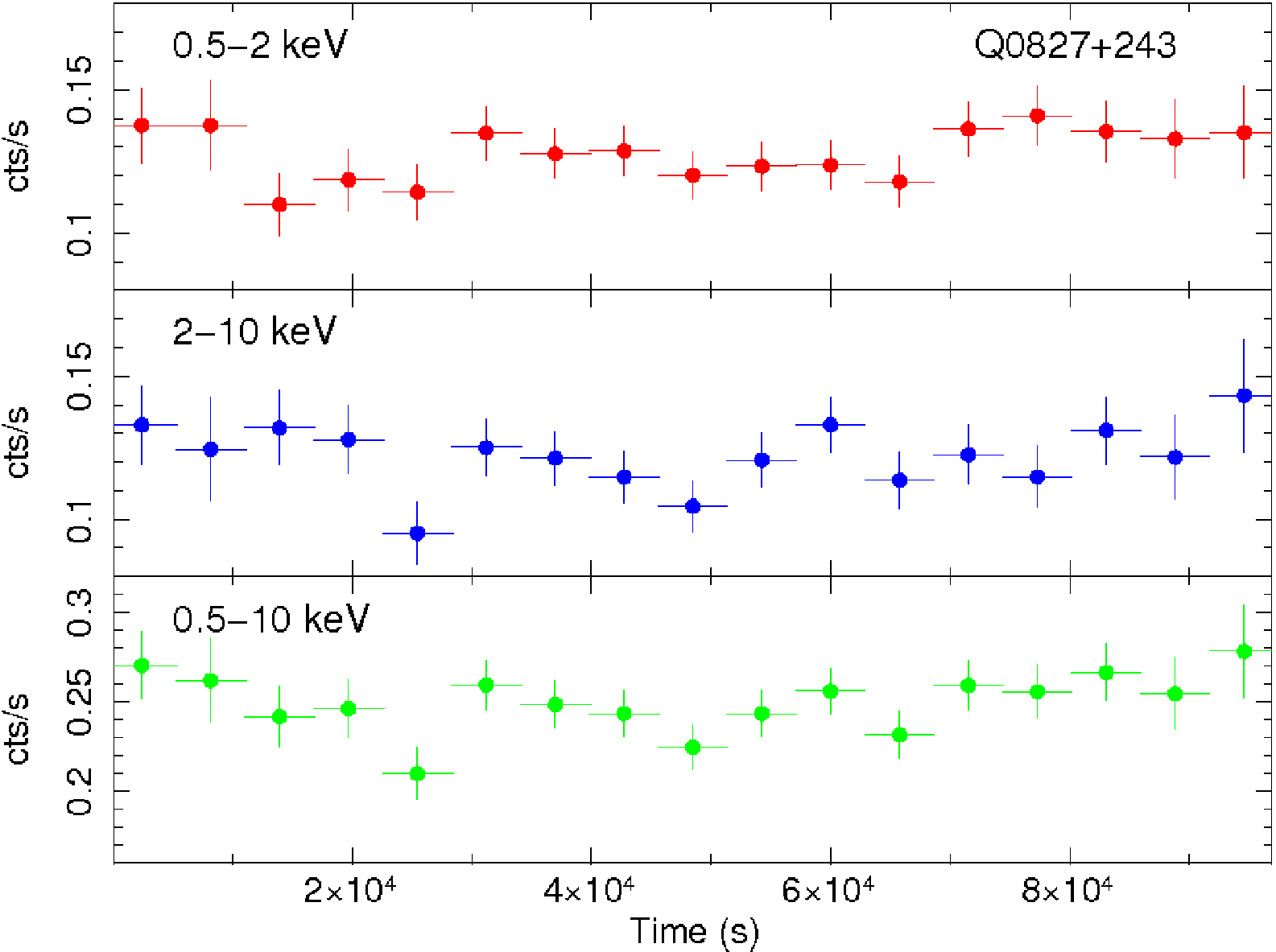}
\includegraphics[angle=0,scale=.3]{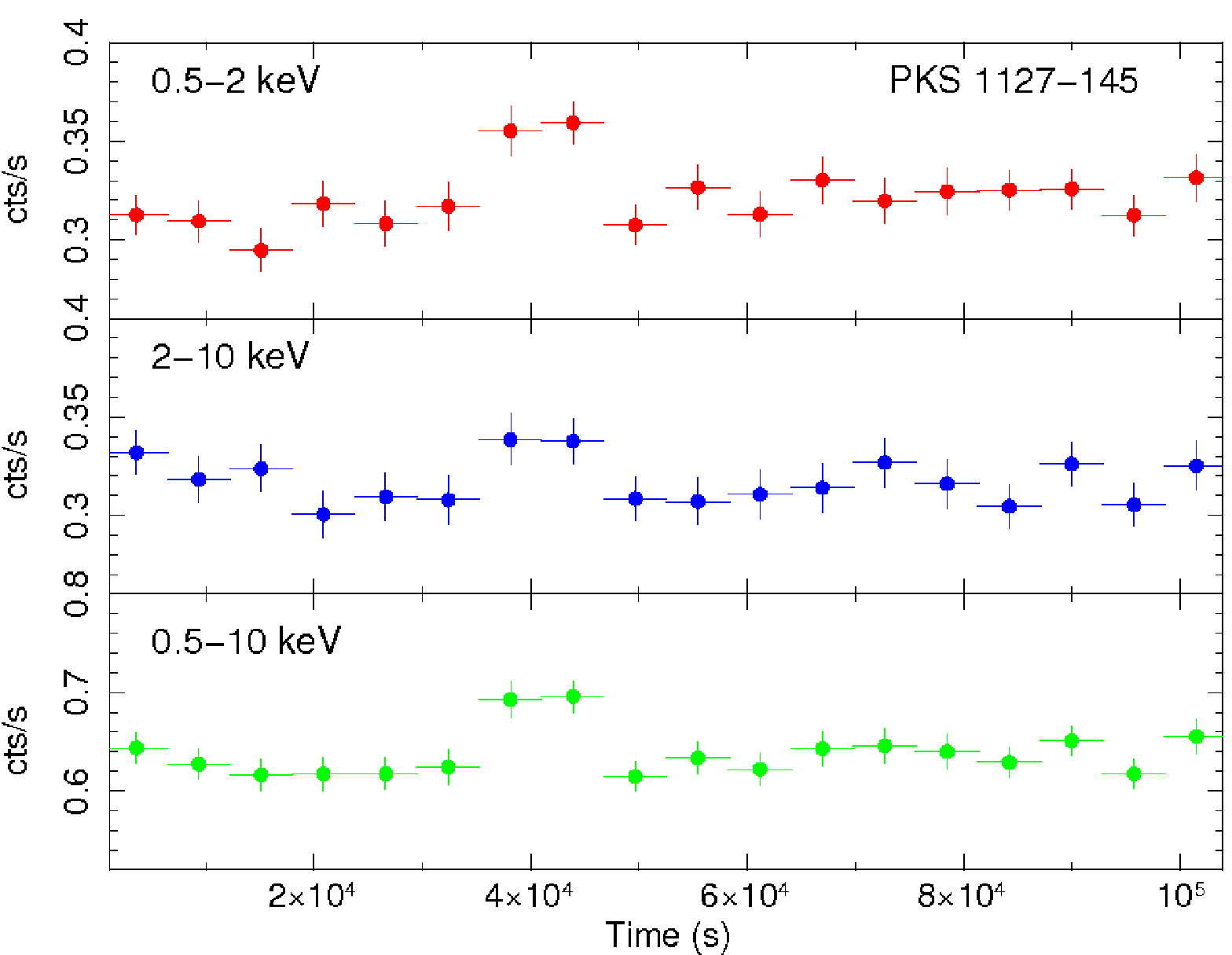}
\includegraphics[angle=0,scale=.3]{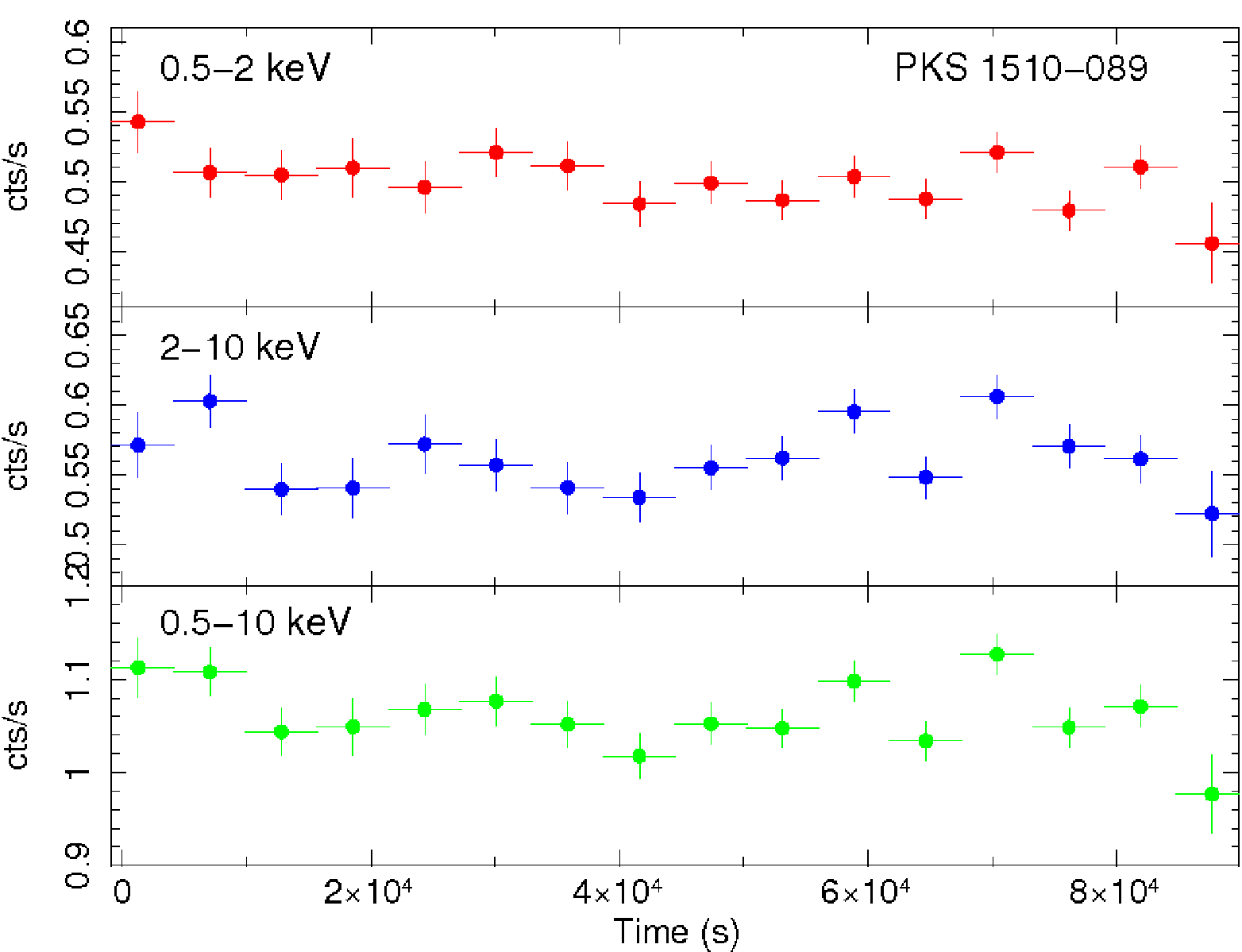}
\includegraphics[angle=0,scale=.3]{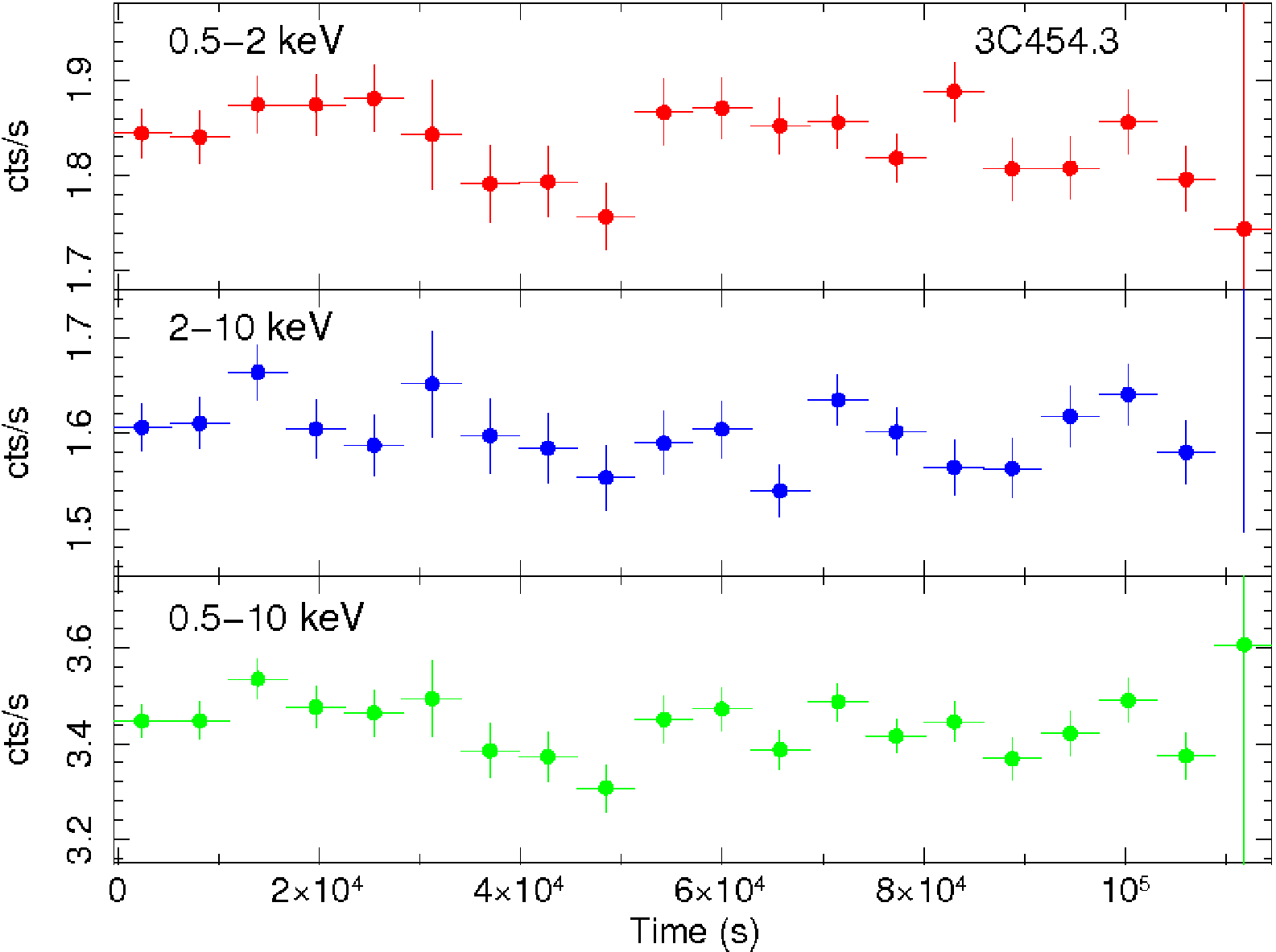}
\caption{Light curves of five FSRQs: $0.5-2$\,keV ({\it upper panels}), $2-10$\,keV 
({\it middle panels}), and $0.5-10$\,keV ({\it bottom panels}). All the light curves were 
binned at $5760$\,s, corresponding to the period of the {\it Suzaku} orbit. 
\label{fig:xis_lc}}
\end{center}
\end{figure}

Figure \ref{fig:xis_lc} shows the count rate variations of the five observed FSRQs.
The summed XISs (XIS0,1,3) light curves are shown separately in different 
energy bands: 0.5$-$2\,keV ({\it upper} panel), 2$-$10\,keV ({\it middle} panel), 
and 0.5$-$10\,keV ({\it bottom} panel), respectively.
Since the count rate variations of the HXD/PIN detector were less clear 
due to limited photon statistics and uncertainty of the modeling of the non-X-ray background, 
we only concentrate on the temporal variability of the XIS data 
below 10\,keV. 
We evaluate the fractional variability by calculating the variability amplitude 
relative to the mean count rate corrected for effects of random errors
(e.g., Edelson et al. 2002): 
$F_{\rm var} = \sqrt{S^2 - \overline{\sigma_{\rm err}}^2}/\overline{x}$, 
where $S^2$ is the total variance of the light curve, $\overline{\sigma_{\rm err}}^2$ 
is the mean error squared and $\overline{x}$ is the mean count rate. 
The variability amplitude in the XIS bands are 
$F_{\rm var} = 0.036\pm0.021$ for 0208$-$512, 
$F_{\rm var} = 0.027\pm0.010$ for 1127$-$145, and 
$F_{\rm var} = 0.025\pm0.008$ for 1510$-$089, 
respectively.  
0827+243 and 3C454.3 show only weak variability, which is not significant.

\subsubsection{Time-averaged spectral analysis}

In the following we report the analysis procedure and results for each object.
The background-subtracted spectra were fitted using XSPEC ver.11.3.2. 
All errors are quoted at the $90\%$ confidence level
for the parameter of interest unless otherwise stated.
All the fits in this paper are restricted to the energy ranges of $0.5-10$\,keV 
(XIS0,3: the FI chips), $0.3-8$\,keV (XIS1: the BI chip), $12-40$\,keV for 
PKS\,1510$-$089, and $12-50$\,keV for 3C\,454.3 (HXD/PIN). We fixed the relative 
normalization of the XISs and HXD/PIN at 1.13, which is carefully determined 
from the XIS calibration using nominal pointings of the Crab Nebula.
Serlemitsos et al. (2007) reported that spectral normalizations 
are slightly different (a few percent) among the CCD sensors based on a 
contemporaneous fit of the Crab spectra. Therefore, we adjusted the 
normalization factor among the three XISs relative to XIS0. 
The results of the spectral fits with a simple absorbed power-law model 
are summarized in Table\,\ref{table:fitresult} (with Galactic absorption) 
and Table\,\ref{table:bestfitresult}.

\begin{table}[ht]
\footnotesize
\caption{Results of the spectral fits to the {\it Suzaku} data using a power-law 
with Galactic absorption}
\begin{center}
\label{table:fitresult}
\begin{tabular}{lccclc}
\tableline
Object      & $N_{\rm H}$\tablenotemark{a}  & $\Gamma$ & $F_{2-10 \rm keV}$\tablenotemark{b} & constant     & $\chi^2_{\rm r}$ \\
            &              &          &                    & (XIS0,1,3,HXD/PIN) &    (dof)        \\
\tableline
0208$-$512 & 3.08 (fixed) & 1.68$\pm$0.03 & 1.37$\pm$0.06 & 1,1.04$\pm$0.05,1.04$\pm$0.05,None & 0.91 (250) \\
0827+243 & 3.62 (fixed) & 1.46$\pm$0.04 & 1.37$\pm$0.07 & 1,0.90$\pm$0.05,1.04$\pm$0.06,None & 0.84 (194) \\ 
1127$-$145 & 3.83 (fixed) & 1.41$\pm$0.02 & 3.45$\pm$0.08 & 1,1.03$\pm$0.03,1.05$\pm$0.03,None & 1.01 (331) \\ 
1510$-$089 & 7.88 (fixed) & 1.37$\pm$0.01 & 6.31$\pm$0.12 & 1,1.00$\pm$0.02,1.02$\pm$0.02,1.13 & 1.06 (407) \\ 
3C\,454.3 & 7.24 (fixed) & 1.58$\pm$0.01 & 16.7$\pm$0.2  & 1,1.04$\pm$0.01,1.02$\pm$0.01,1.13 & 1.00 (1090) \\ 
\tableline
\end{tabular}
\tablenotetext{}{Errors correspond to 90\% confidence level.} 
\tablenotetext{a}{Fixed value indicates the Galactic absorption column density in units of $10^{20}$\,cm$^{-2}$.}
\tablenotetext{b}{Flux in units of $10^{-12}$\,erg\,cm$^{-2}$\,s$^{-1}$.}
\end{center}
\end{table}

\begin{table}[ht]
\footnotesize
\caption{Results of the spectral fits to the {\it Suzaku} data with best fit models}
\begin{center}
\label{table:bestfitresult}
\begin{tabular}{lccccccc}
\tableline
Object      & Model\tablenotemark{a} & $N_{\rm H}$ & $\Gamma_{\rm hi}$\tablenotemark{b} & $\Gamma_{\rm low}$\tablenotemark{c} & $F_{2-10 \rm keV}$\tablenotemark{d} & kT     & $\chi^2_{\rm r}$ \\
            &       &             &                   &                    &                                     & (keV)  &                  \\
\tableline
0208$-$512 & PL    & 3.08 (fixed) & 1.68$\pm$0.03 & - & 1.37$\pm$0.06 & - & 0.91 (250) \\
0827+243 & PL    & 3.62 (fixed) & 1.46$\pm$0.04 & - & 1.37$\pm$0.07 & - & 0.84 (194) \\ 
1127$-$145 & PL    & 10.8$^{+1.6}_{-1.5}$ & 1.52$\pm$0.03 & - & 3.36$\pm$0.08 & - & 0.82 (330) \\ 
1510$-$089 & PL+BB & 7.88 (fixed) & 1.32$\pm$0.03 & - & 6.42$\pm$0.13 & 0.15$\pm$0.03 & 0.97 (405) \\ 
         & PL+PL & 7.88 (fixed) & 1.26$^{+0.06}_{-0.12}$ & 2.85$^{+0.88}_{-0.40}$ & 6.30$^{+0.18}_{-0.74}$ & - & 0.96 (405) \\  
3C\,454.3 & PL   & 9.07$^{+0.58}_{-0.57}$ & 1.62$\pm$0.01 & - & 16.6$\pm$0.2 & - & 0.97 (1089) \\ 
\tableline
\end{tabular}
\tablenotetext{a}{Spectral fitting models. PL, power-law function; PL+PL, double power-law function; PL+BB, power-law + blackbody model.}
\tablenotetext{b}{Differential spectral photon index.}
\tablenotetext{c}{Differential spectral photon index at the low-energy X-ray band, when fitted with a double power-law function.}
\tablenotetext{d}{Flux in units of $10^{-12}$\,erg\,cm$^{-2}$\,s$^{-1}$.}
\end{center}
\end{table}

\begin{figure}[ht]
\begin{center}
\includegraphics[angle=0,scale=.25]{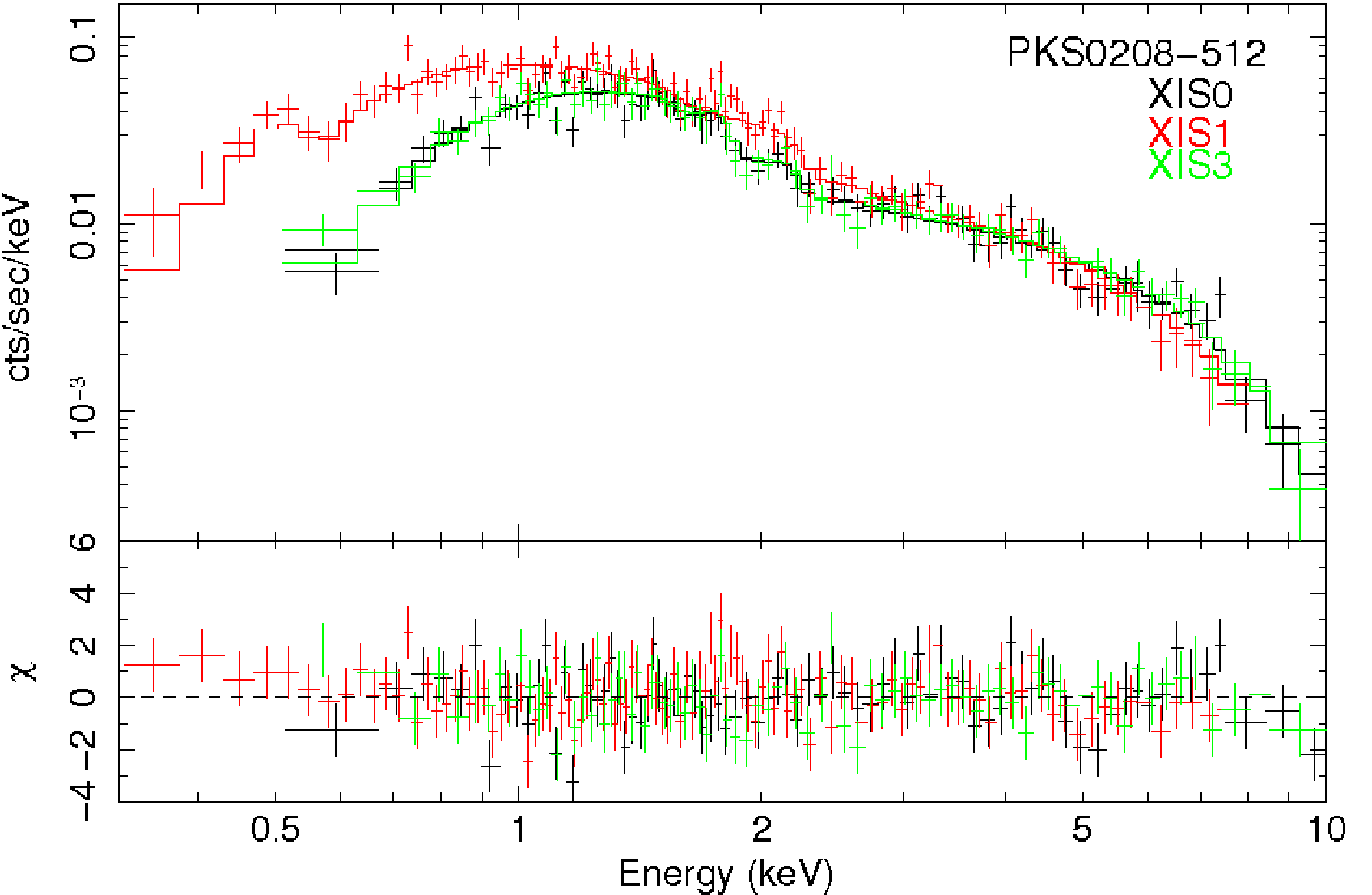}
\includegraphics[angle=0,scale=.25]{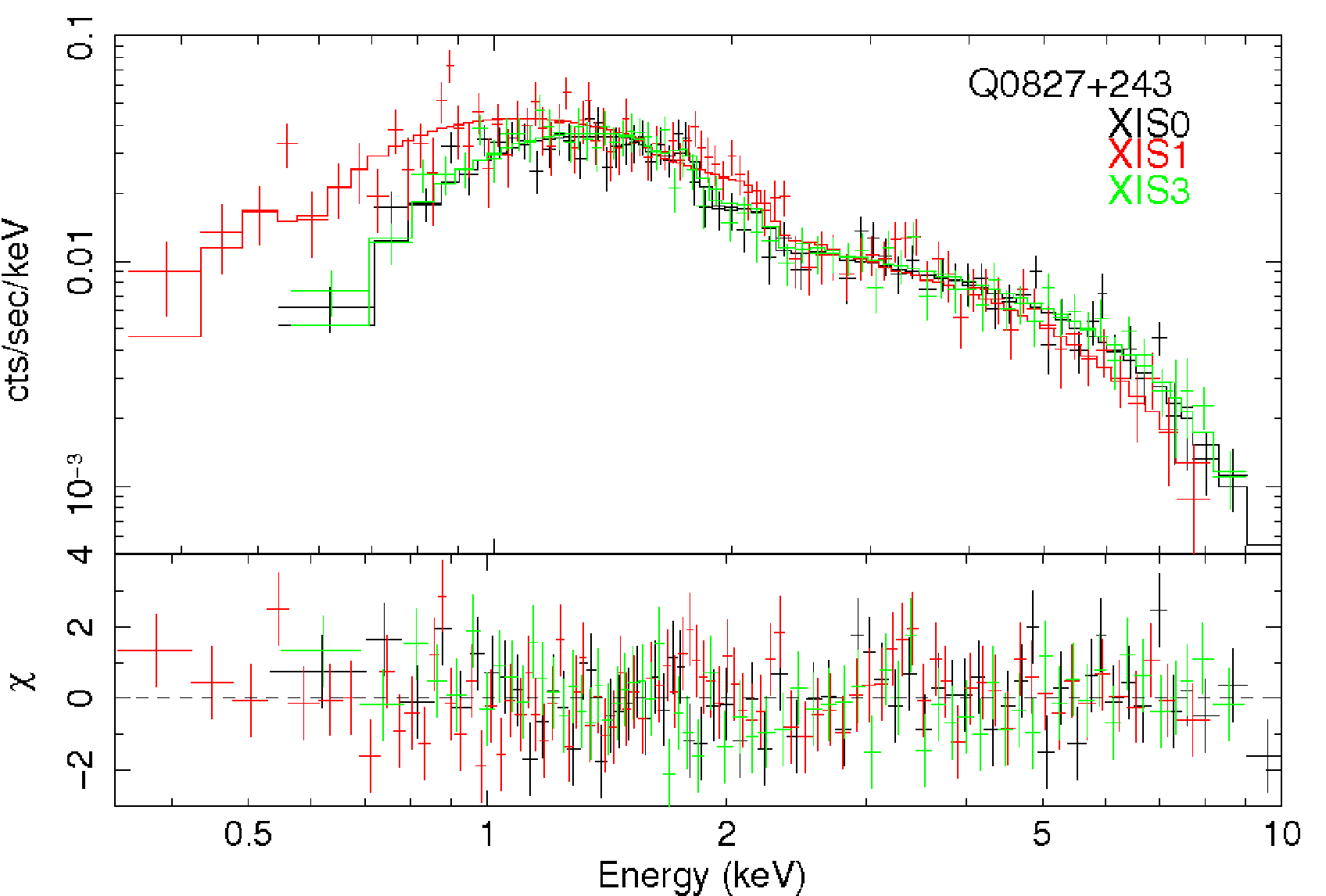}
\includegraphics[angle=0,scale=.25]{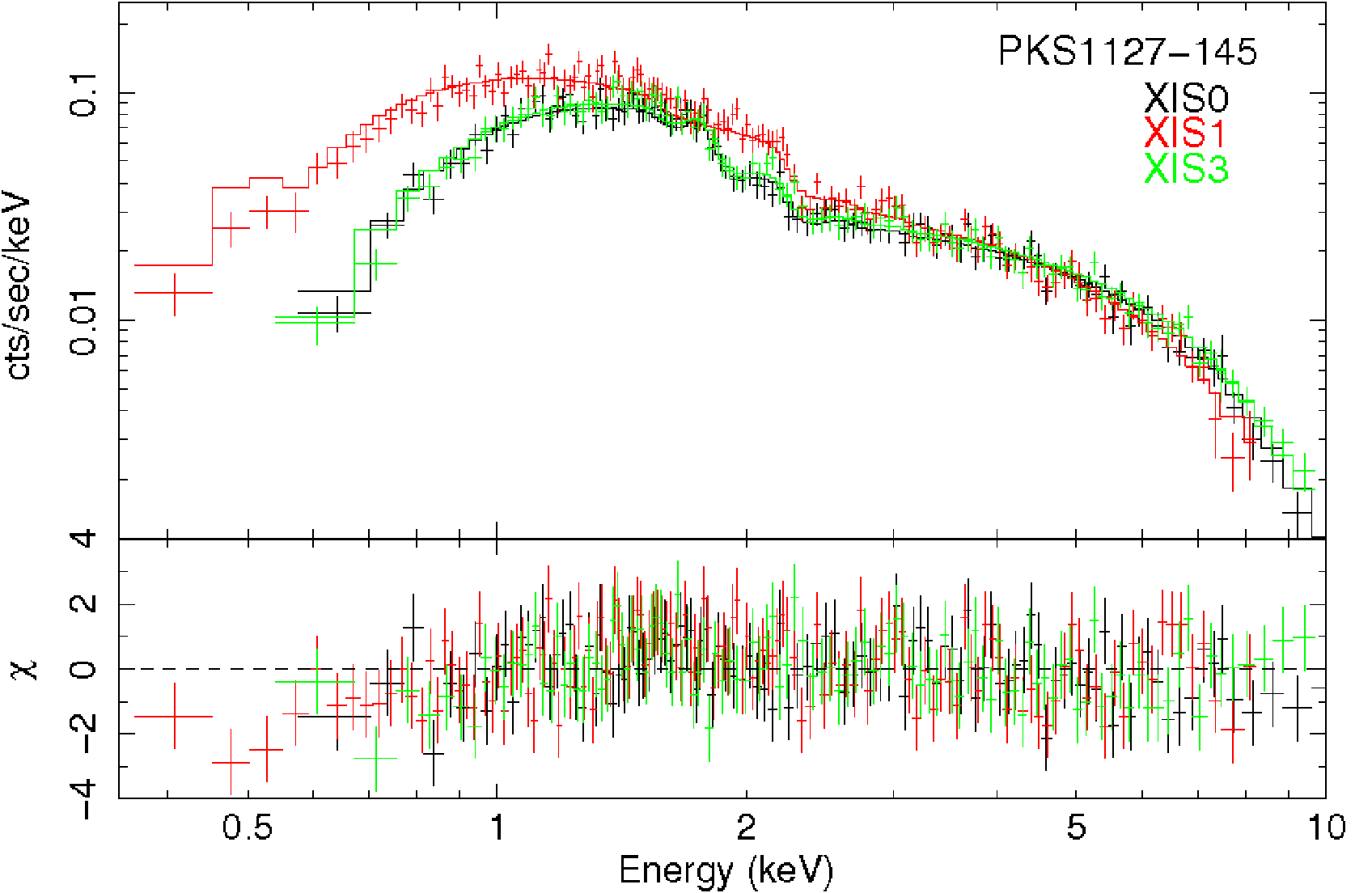}
\includegraphics[angle=0,scale=.25]{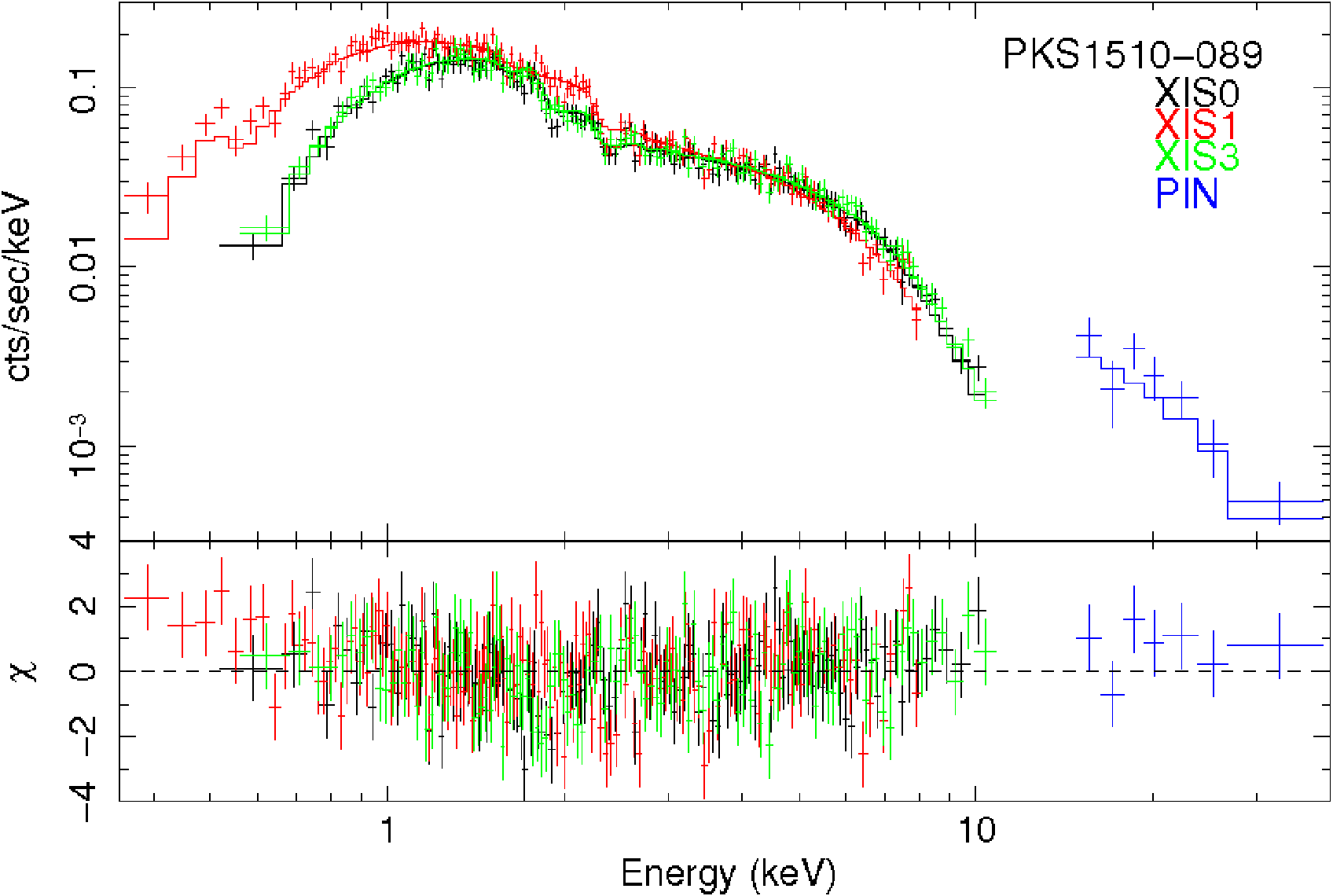}
\includegraphics[angle=0,scale=.25]{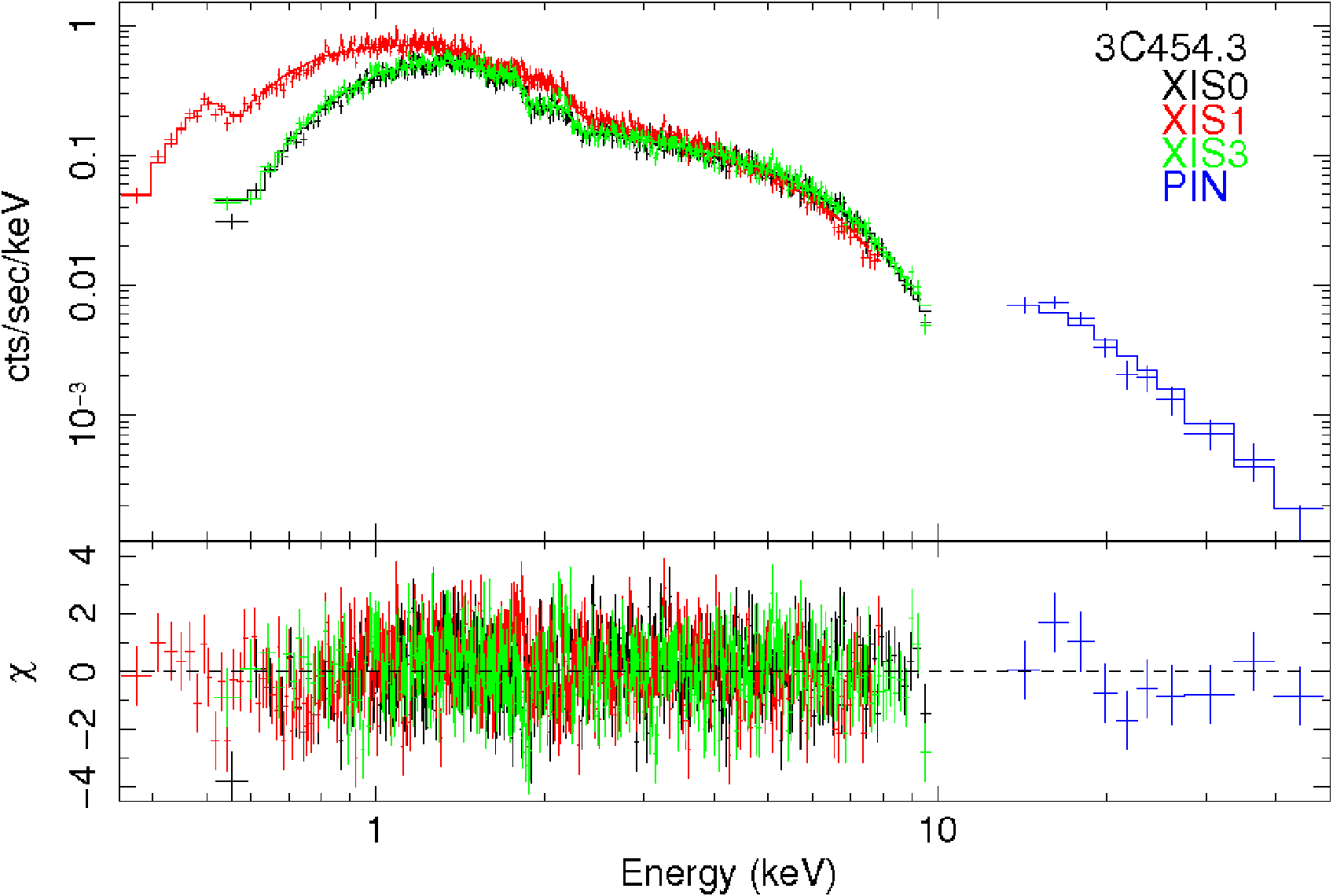}
\caption{{\it Suzaku} spectra of five FSRQs: the top panel shows the data,
plotted against a power-law model with the Galactic absorption. 
The bottom panel shows the residuals for the power-law fit. 
For 0208$-$512 and 1510$-$089, the data below 1 keV are in excess to the model. 
On the other hand, for 1127$-$145, the residuals show a substantial deficit 
of photons at low energies. For 3C454.3, some scatter around 1 keV in the residual panel is seen.\label{fig:QHBspec}}
\end{center}
\end{figure}

\begin{figure}[ht]
\begin{center}
\includegraphics[angle=0,scale=.25]{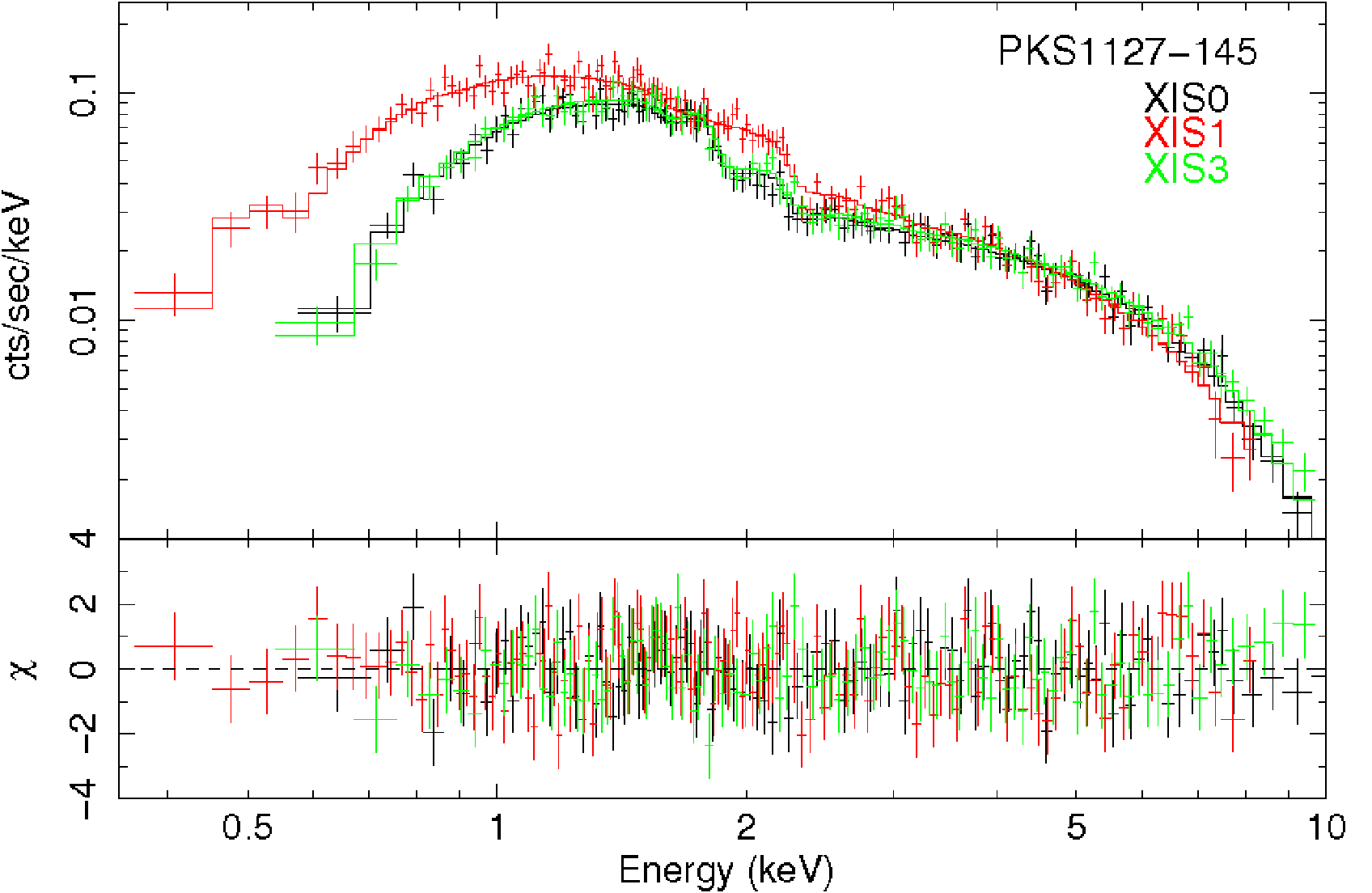}
\includegraphics[angle=0,scale=.25]{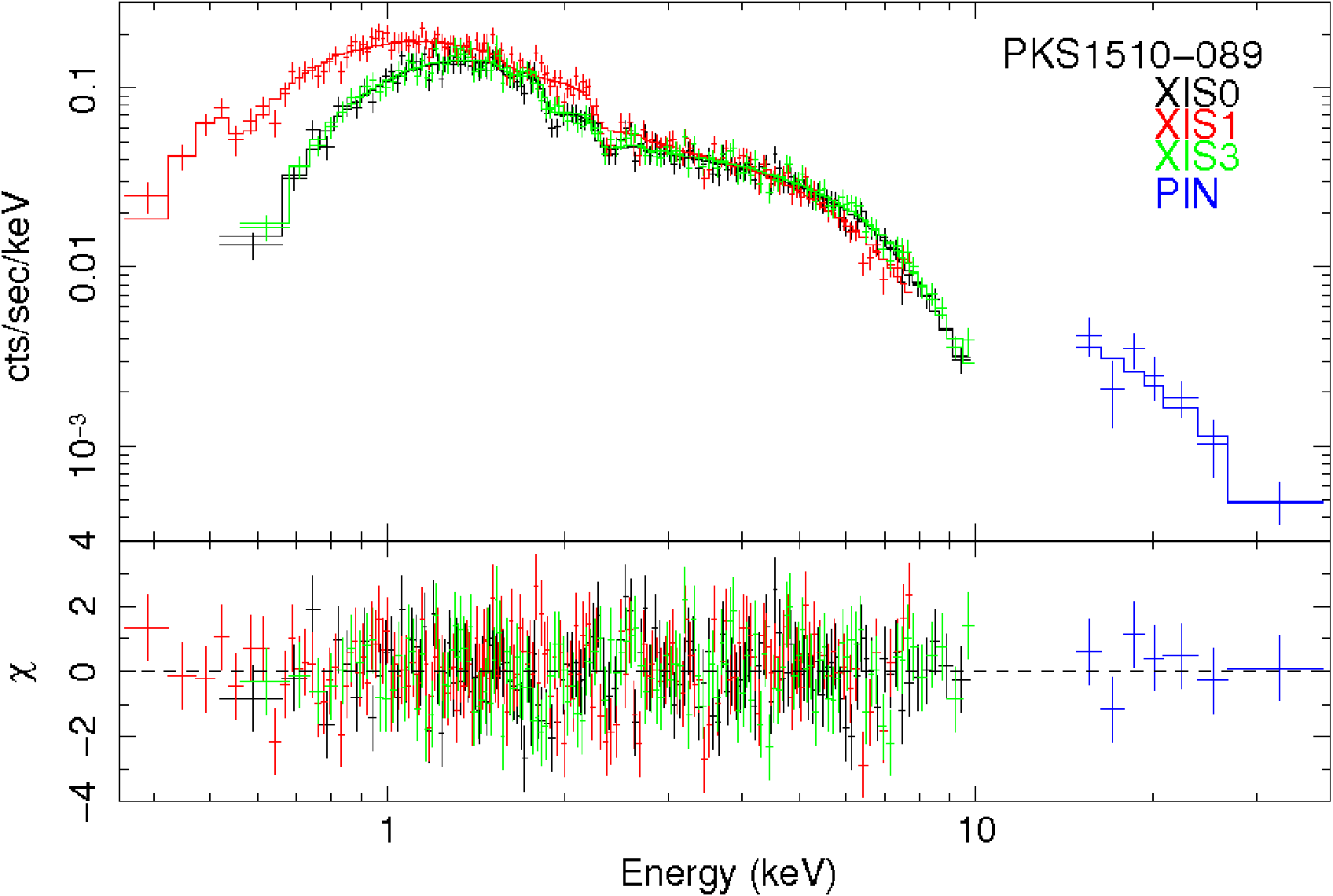}
\includegraphics[angle=0,scale=.25]{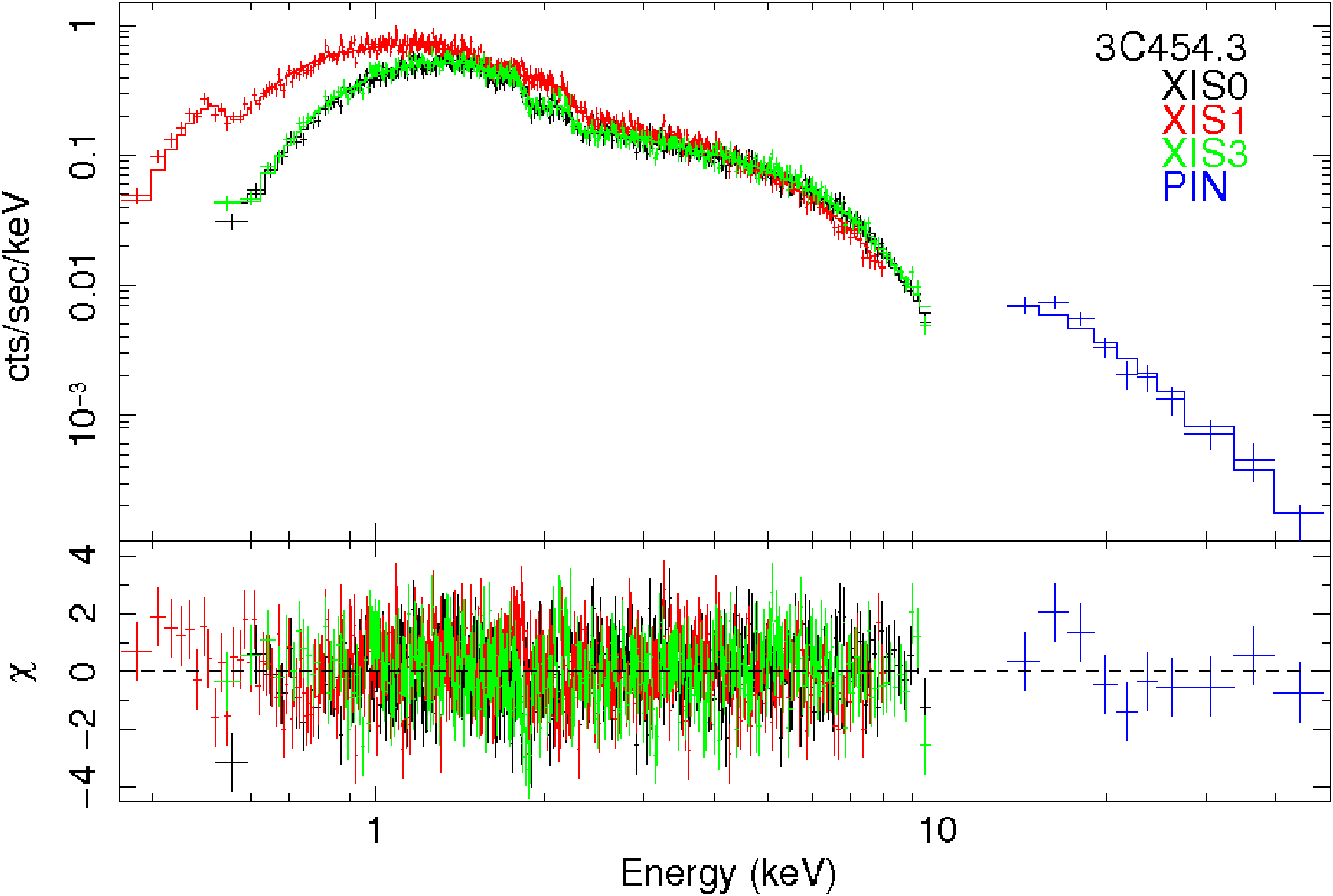}
\caption{Best fit {\it Suzaku} spectrum of PKS\,1127$-$145, PKS\,1510$-$089 and 3C454.3. 
The top panel shows the data, plotted against an absorbed power-law model. 
The bottom panel shows the residuals to the power-law fit. \label{fig:1127spec}}
\end{center}
\end{figure}

\subsubsubsection{PKS\,0208$-$512}

The time averaged, background subtracted three XIS spectra of PKS\,0208$-$512, 
when fitted jointly, are well described by a single absorbed power-law model, and the absorption 
column is consistent with the Galactic value $N_{\rm H} = 3.08\times10^{20}$\,cm$^{-2}$ 
(Dickey \& Lockman 1990). We obtained the best fit photon index $\Gamma = 1.68\pm0.02$ 
and the $2-10$\,keV flux $F_{2-10 \rm keV} = (1.37\pm0.03) \times 
10^{-12}$\,erg\,cm$^{-2}$\,s$^{-1}$ with a chi-squared value of $0.91$ for $250$ dof.  
Figure\,\ref{fig:QHBspec} shows the spectra obtained with the XISs with residuals 
plotted against the best-fit power-law model with Galactic absorption. 
Although statistically acceptable, we notice that the residuals of the fits show 
moderate excess feature at low energies, below $1$\,keV. 

\begin{figure}[ht]
\begin{center}
\includegraphics[angle=0,scale=.25]{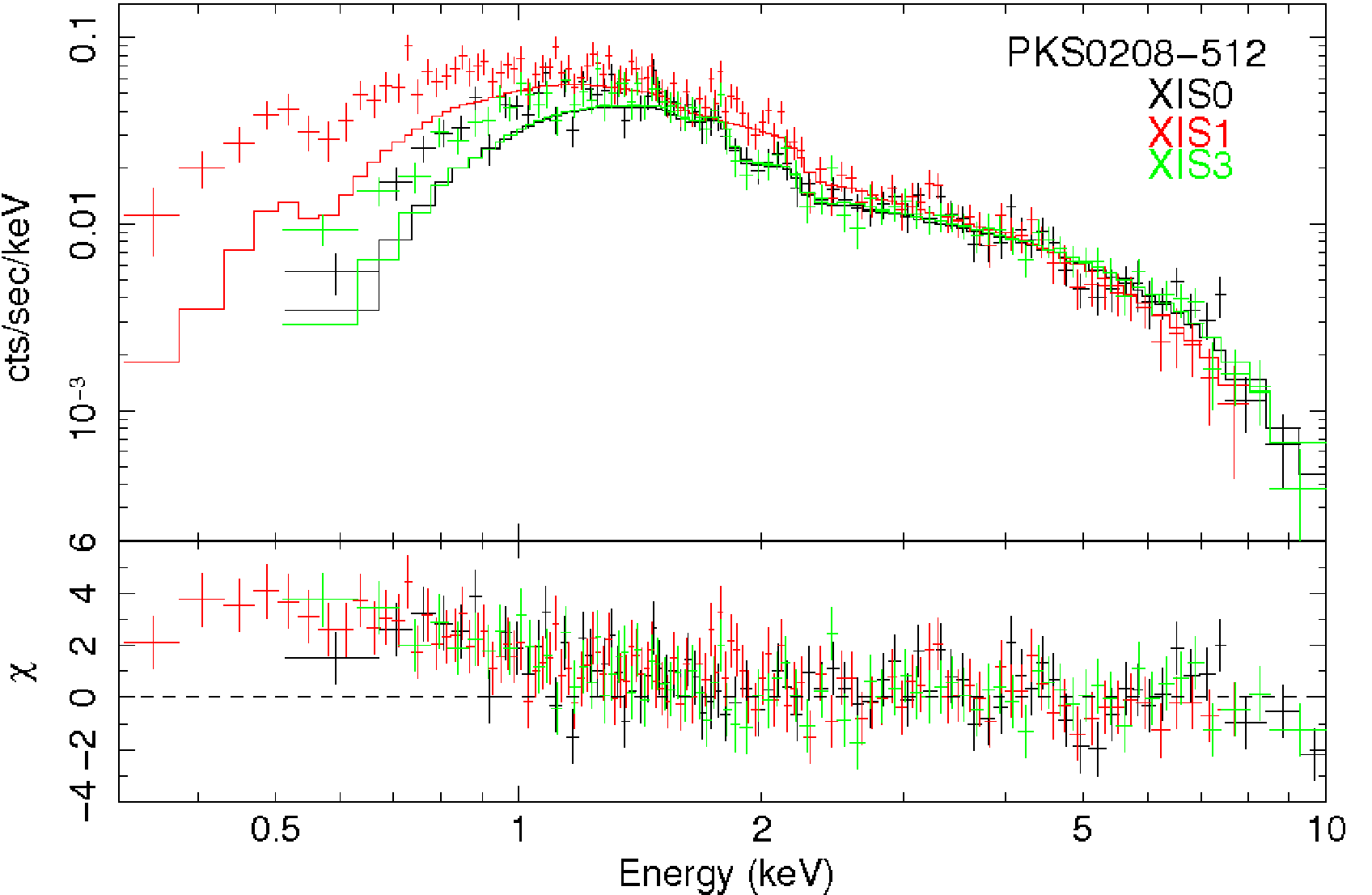}
\caption{{\it Suzaku} spectrum of PKS 0208$-$512, with residuals assuming a column density of 
$N_{\rm H} \sim 1.67\times10^{21}$\,cm$^{-2}$.
Deviations due to soft excess emission can be clearly seen. \label{fig:0208spec}}
\end{center}
\end{figure}

In the previous observation with BeppoSAX during a high flux state (Tavecchio et al. 
2002), $F_{2-10 \rm keV} \sim 4.7 \times 10^{-12}$\,erg\,cm$^{-2}$\,s$^{-1}$ - 
which is a factor of three larger than in our {\it Suzaku} observations - 
the X-ray spectrum is well described by a power-law with photon index 
$\Gamma\sim1.7$, similar to the {\it Suzaku} result. However, Tavecchio et al. 
reported that the spectrum was heavily absorbed below $1$\,keV, indicating a 
column density of $N_{\rm H} = 1.67\times10^{21}$\,cm$^{-2}$. Figure\,\ref{fig:0208spec} 
shows the {\it Suzaku} spectrum with residuals assuming such an 
increased value of $N_{\rm H}$. The residuals indicate significant 
soft excess emission below $1$\,keV, if $N_{\rm H}$ is the same as found in the 
previous BeppoSAX observation. 

The variable soft X-ray emission of PKS\,0208$-$512 may indicate 
that the convex spectrum observed by BeppoSAX reflects an intrinsic IC 
continuum shape, while the soft excess observed by {\it Suzaku} reflects the presence of 
an additional spectral component which becomes prominent when the source gets fainter 
(see Tavecchio et al. 2007; Kataoka et al. 2008). Therefore, to model in more detail the 
observed X-ray spectrum, we first considered a double power-law fit 
(PL + PL) in which the soft X-ray excess is represented by a steep power-law component. 
The absorption column is fixed at  $N_{\rm H} = 1.67\times10^{21}$\,cm$^{-2}$, as 
given by Tavecchio et al. (2002). We obtained the photon indices $\Gamma_1=4.98\pm0.30$ 
and $\Gamma_2=1.71_{-0.04}^{+0.02}$. This provides an acceptable fit, with $\chi^2_{\rm r}$/dof 
= 0.90/248. We also considered an alternative fit consisting of a power-law 
function and a blackbody component. This model also gives a similarly good representation 
of the data with $\chi^2_{\rm r}$/dof = 0.90/248, implying $\Gamma=1.78\pm0.03$ 
and the temperature of the introduced thermal component of $kT = 0.092\pm0.003$\,keV. 
Both fits appear to be as good as a single power-law with free absorption, 
and do not improve the goodness of fit. 

\subsubsubsection{Q\,0827+243}

The time averaged spectra of Q\,0827+243 collected with the XISs are well fitted by an absorbed 
power-law model with a photon index $\Gamma = 1.46\pm0.02$ ($\chi^2_{\rm r} = 0.84$ for 194 dof). 
The absorption column is consistent with the Galactic value of 
$N_{\rm H} = 3.62 \times 10^{20}$\,cm$^{-2}$ (Dickey \& Lockman 1990), and the flux 
over $2-10$\,keV is $F_{2-10 \rm keV} = (1.37\pm0.04)\times10^{-12}$\,erg\,cm$^{-2}$\,s$^{-1}$. 
As shown in Figure\,\ref{fig:QHBspec}, there is no evidence for any additional 
spectral feature in the soft band. This result is in good agreement with previous 
{\it Chandra} observations of the core (Jorstad \& Marscher 2004), revealing that the X-ray continuum
is well described by a power-law model ($\Gamma \sim 1.4$) with Galactic absorption.

\subsubsubsection{PKS\,1127$-$145}

We first fitted the XISs spectra with a single power-law model with a Galactic absorption 
of $N_{\rm H} = 3.83\times10^{20}$\,cm$^{-2}$ (Murphy et al. 1996). 
We obtained the photon index of $\Gamma = 1.41\pm0.01$ $\chi^2_{\rm r} = 1.01$ for 331 dof), 
but the residuals show a substantial deficit of photons at low energies (Figure\,\ref{fig:QHBspec}). 
To investigate this deficit in more detail, we fitted the spectra with a single power-law 
and a free absorption model. This model represents well the spectra with the best chi-squared value 
of 0.82 for 330 dof (Figure\,\ref{fig:1127spec}),
indicating that the column density is higher than the Galactic value at the $99.9\%$ 
confidence level. For this model the photon index is $\Gamma = 1.51\pm0.02$ and the 
unabsorbed X-ray flux is $F_{2-10 \rm keV} = (3.36\pm0.05)\times10^{-12}$\,erg\,cm$^{-2}$\,s$^{-1}$. 
The best-fit column density is $N_{\rm H} = (1.08\pm0.09) \times 10^{21}$\,cm$^{-2}$, 
which is similar to the one found in previous {\it Chandra} and XMM-{\it Newton} 
observations ($N_{\rm H} \sim 1.2 \times 10^{21}$\,cm$^{-2}$) during a high state 
with $F_{2-10 \rm keV} \sim 6\times10^{-12}$\,erg\,cm$^{-2}$\,s$^{-1}$ 
(Bechtold et al. 2001; Foschini et al. 2006). 
We note that the Galactic absorption and a broken power-law model also well
represents the spectra with $\chi^2_{\rm r}$/dof of 0.81/329.
In this model, the spectrum below $E_{\rm brk} = 1.50\pm0.08$\,keV is rather hard
($\Gamma_1 = 1.10\pm0.05$), and the high energy photon index is $\Gamma_2 = 1.50\pm0.02$.

\subsubsubsection{PKS\,1510$-$089}

Figure\,\ref{fig:QHBspec} shows the XISs and HXD/PIN spectra of PKS\,1510$-$089 
(including residuals), plotted against the best-fit power-law model with Galactic 
absorption, using the overall X-ray data between $0.3$ and $40$\,keV.  
The best fit photon index is $\Gamma = 1.37\pm0.01$ and the unabsorbed X-ray
flux is $F_{2-10 \rm keV} = (6.31\pm0.07)\times10^{-12}$\,erg\,cm$^{-2}$\,s$^{-1}$. 
However, this model did not represent the spectra well yielding a chi-squared value of 
1.06 for 407 dof. The residuals indicate some excess emission at low energies. 

To represent the observed X-ray spectra, we tried the same analysis as for PKS\,0208$-$512.
We first fitted the data by a double power-law model with Galactic absorption. 
We obtained the photon indices $\Gamma_1=2.84_{-0.47}^{+0.50}$ and
$\Gamma_2=1.26_{-0.06}^{+0.04}$. This provides an acceptable fit, with $\chi^2_{\rm r}$/dof = 0.96/405. 
The improvement of the chi-squared statistic is significant 
at more than the $99.9\%$ confidence level when compared to the single power-law 
model. Next, we considered an alternative fit consisting of a power-law function 
and a blackbody component. This model also gives a good representation of the 
data, with $\chi^2_{\rm r}$ of 0.97 for 405 dof, indicating that the photon index 
is $\Gamma=1.32\pm0.02$ and the temperature of the introduced thermal component is $kT=0.15\pm0.02$\,keV. 
This result is consistent with previous {\it Suzaku} ($\Gamma=1.24\pm0.01$; 
Kataoka et al. 2008) and BeppoSAX observations ($\Gamma=1.39\pm0.08$; 
Tavecchio et al. 2000).

\subsubsubsection{3C\,454.3}

We first fitted the XISs and PIN spectra with a single power-law model with a Galactic absorption 
of $N_{\rm H} = 7.24\times10^{20}$\,cm$^{-2}$ (Murphy et al. 1996). 
We obtained the photon index of $\Gamma = 1.41\pm0.01$ ($\chi^2_{\rm r} = 1.00$ for 1090 dof), 
but the residuals show some scatter around 1 keV. 
To investigate this scatter in more detail, we fitted the spectra 
with a single power-law and a free absorption model. This model represents well 
the spectra with the best chi-squared value of 0.97 for 1089 dof, indicating that the column density 
is higher than the Galactic value at the $99.9\%$ confidence level. For this model 
the photon index is $\Gamma = 1.51\pm0.02$ and the unabsorbed X-ray flux is 
$F_{2-10 \rm keV} = (3.36\pm0.05)\times10^{-12}$\,erg\,cm$^{-2}$\,s$^{-1}$. 

In the case of 3C\,454.3 we obtained the best fit to the XISs and HXD/PIN 
spectra assuming an absorbed power-law model with a photon index of $\Gamma = 1.61\pm0.01$ 
and a column density of $N_{\rm H} = (9.07\pm0.35) \times 10^{20}$\,cm$^{-2}$,
which is larger than the Galactic value at the $99.9\%$ confidence level.
The unabsorbed $2-10$\,keV flux is $F_{2-10 \rm keV} = (1.66\pm0.01)
\times10^{-11}$\,erg\,cm$^{-2}$\,s$^{-1}$ (Figure \ref{fig:QHBspec}).
The spectra can be fitted with both the Galactic absorption and a broken power-law model
as well as the above model ($\chi^2_{\rm r}$/dof of 0.96/1088). In the former case,
the photon indices are $\Gamma_1 = 1.47_{-0.03}^{+0.01}$ and $\Gamma_2 = 1.61\pm0.01$,
while the break energy is $E_{\rm brk} = 1.29_{-0.11}^{+0.08}$ keV.
In addition, we reanalyzed the previous {\it Suzaku} data collected in 
December 2007 during the high state (Donnarumma et al. 2010). 
The time averaged XISs and HXD/PIN spectra was well described by a single 
absorbed power-law model with $\Gamma = 1.64\pm0.01$, implying the flux 
$F_{2-10 \rm keV} = (3.09\pm0.02)\times10^{-11}$\,erg\,cm$^{-2}$\,s$^{-1}$,  
which is larger by a factor of two than the one found in our 2008 observations.
The absorption column also shows a higher value of $N_{\rm H} = (1.07\pm0.03) 
\times 10^{21}$\,cm$^{-2}$. 

\begin{figure}[ht]
\begin{center}
\includegraphics[angle=0,scale=.25]{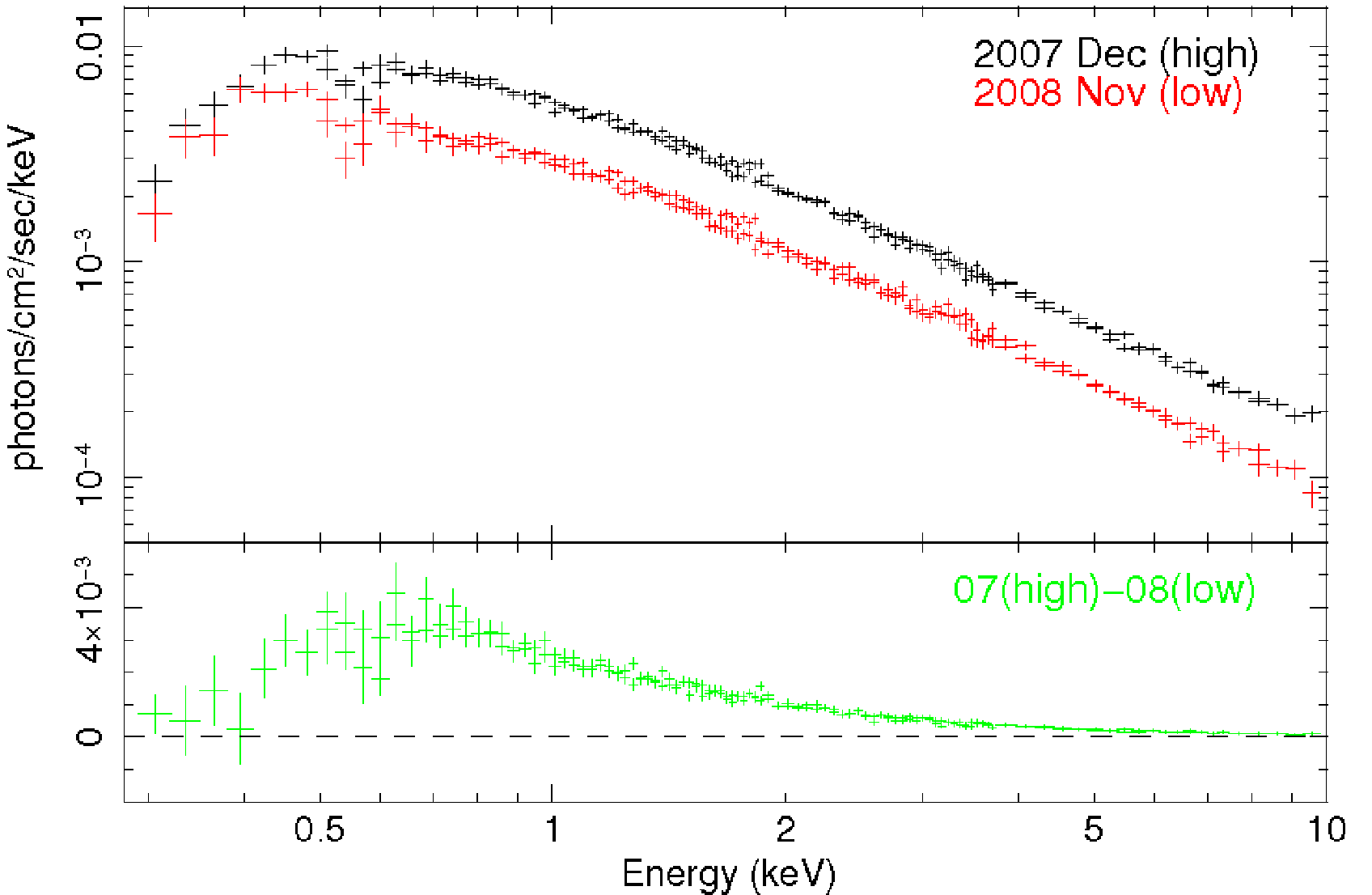}
\caption{Unfolded spectra of 3C\,454.3 obtained in 2007 (high) and 2008 (low),
respectively. The bottom panel shows the residuals by subtracting the spectra 
in the high state from those in the low state. \label{fig:3C454spec}}
\end{center}
\end{figure}

Figure\,\ref{fig:3C454spec} shows the unfolded spectra obtained in 2007 (high state) 
and 2008 (this work; low state). 
The bottom panel shows the residuals by subtracting the spectra in the high state 
from those in the low state. The excess emission at low energies is clearly 
visible in the residuals. 

The previous X-ray observations of 3C\,454.3 often indicated some additional
absorption in excess to the Galactic value. For example, Villata et al. (2006) 
reported $N_{\rm H}=(1.34\pm0.05)\times10^{21}$\,cm$^{-2}$ in the {\it Chandra} 
data collected in May 2005, during the outburst phase ($F_{2-8 \rm keV} \sim 8.4 
\times 10^{-11}$\,erg\,cm$^{-2}$\,s$^{-1}$, which is $\sim 5$ times higher than in 
our observation). An even higher hydrogen column density was found by Giommi et al. (2006), 
when fitting the April-May 2005 data taken by the {\it Swift} XRT ($N_{\rm H} \sim 2-3 
\times10^{21}$\,cm$^{-2}$), and by Atari et al. (2007, 2008), using the July and December 
2006, and May 2007 data taken by XMM-{\it Newton}. Assuming that the intrinsic absorption 
in 3C\,454.3 is the same as reported in Villata et al. (2006), we fit our {\it Suzaku}
data first by a double power-law function, obtaining $\Gamma_1=3.66_{-0.33}^{+0.35}$ and
$\Gamma_2=1.61\pm0.02$. This provides an acceptable fit, with $\chi^2_{\rm r}$/dof = 0.97/1088.
Next we consider an alternative fit consisting of a power-law and a blackbody 
component. This model gives a good representation of the data, with $\chi^2_{\rm r}$/dof = 1.00/1088, 
a photon index of $\Gamma=1.65\pm0.01$, and a temperature 
of $kT=0.105\pm0.004$\,keV. However, the fitting results do not improve the goodness 
of fit compared with the single power-law model with free absorption.

\subsubsection{Time-resolved spectral analysis}

\begin{figure}[ht]
\begin{center}
\includegraphics[angle=0,scale=.3]{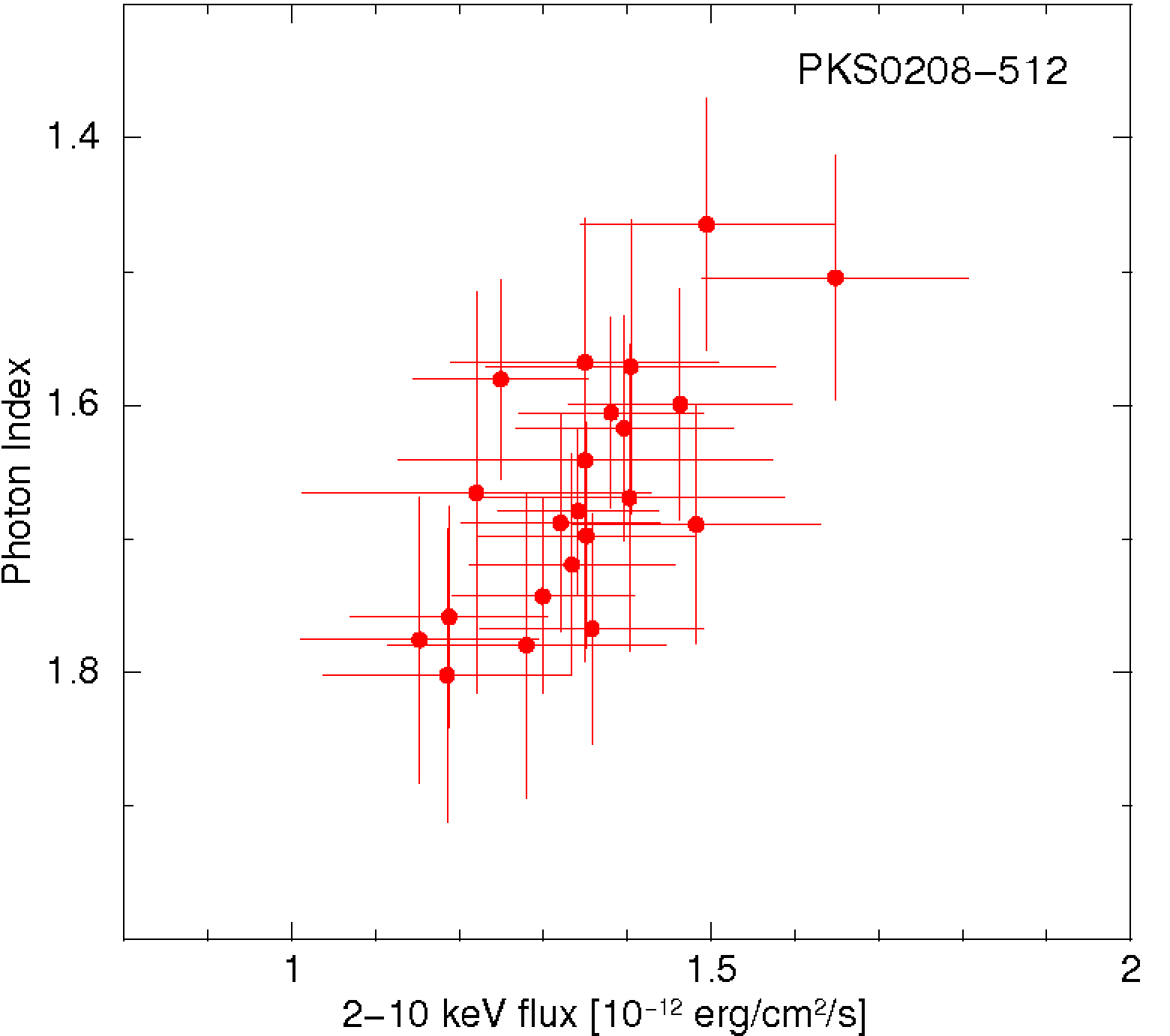}
\includegraphics[angle=0,scale=.3]{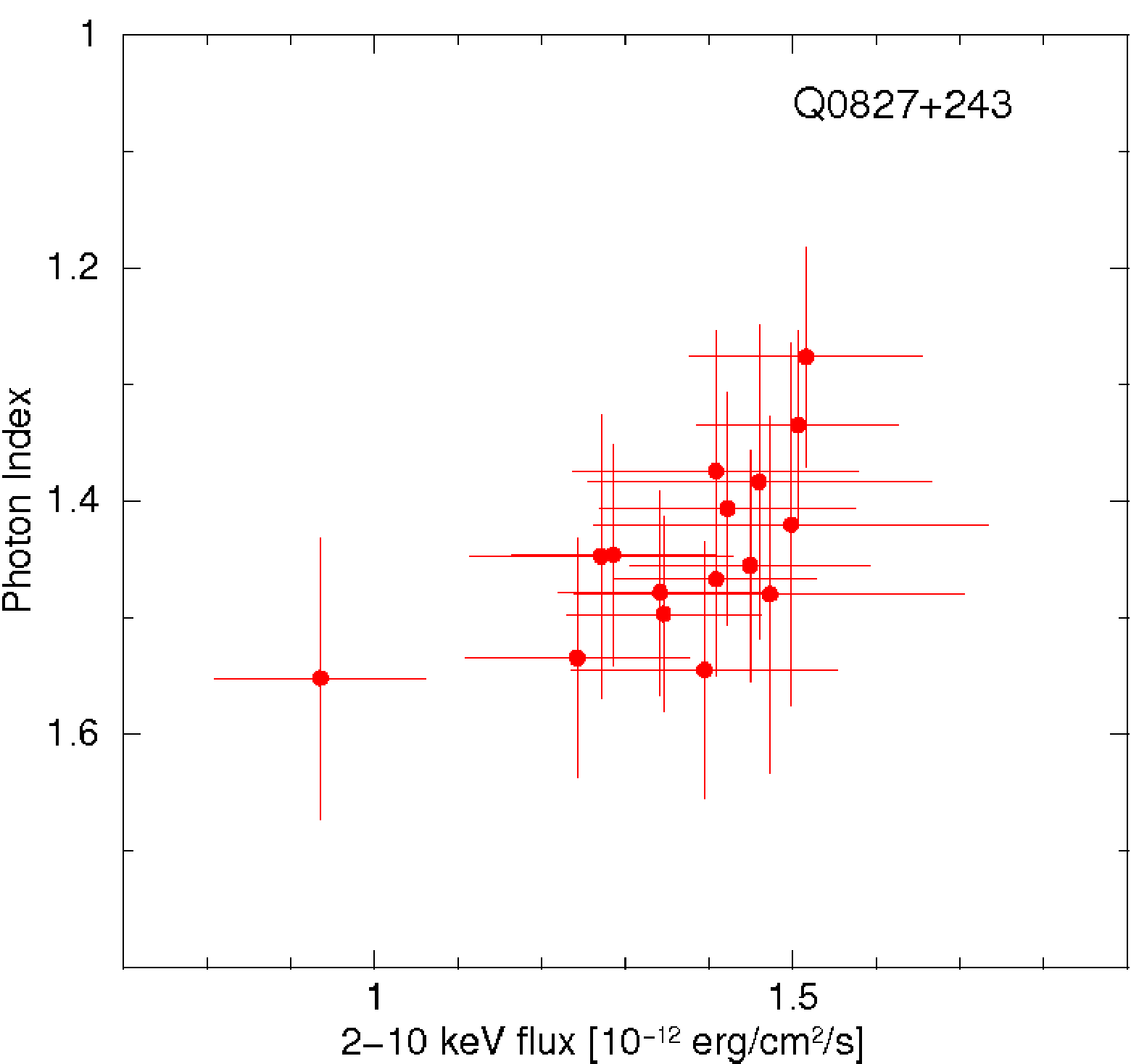}
\includegraphics[angle=0,scale=.3]{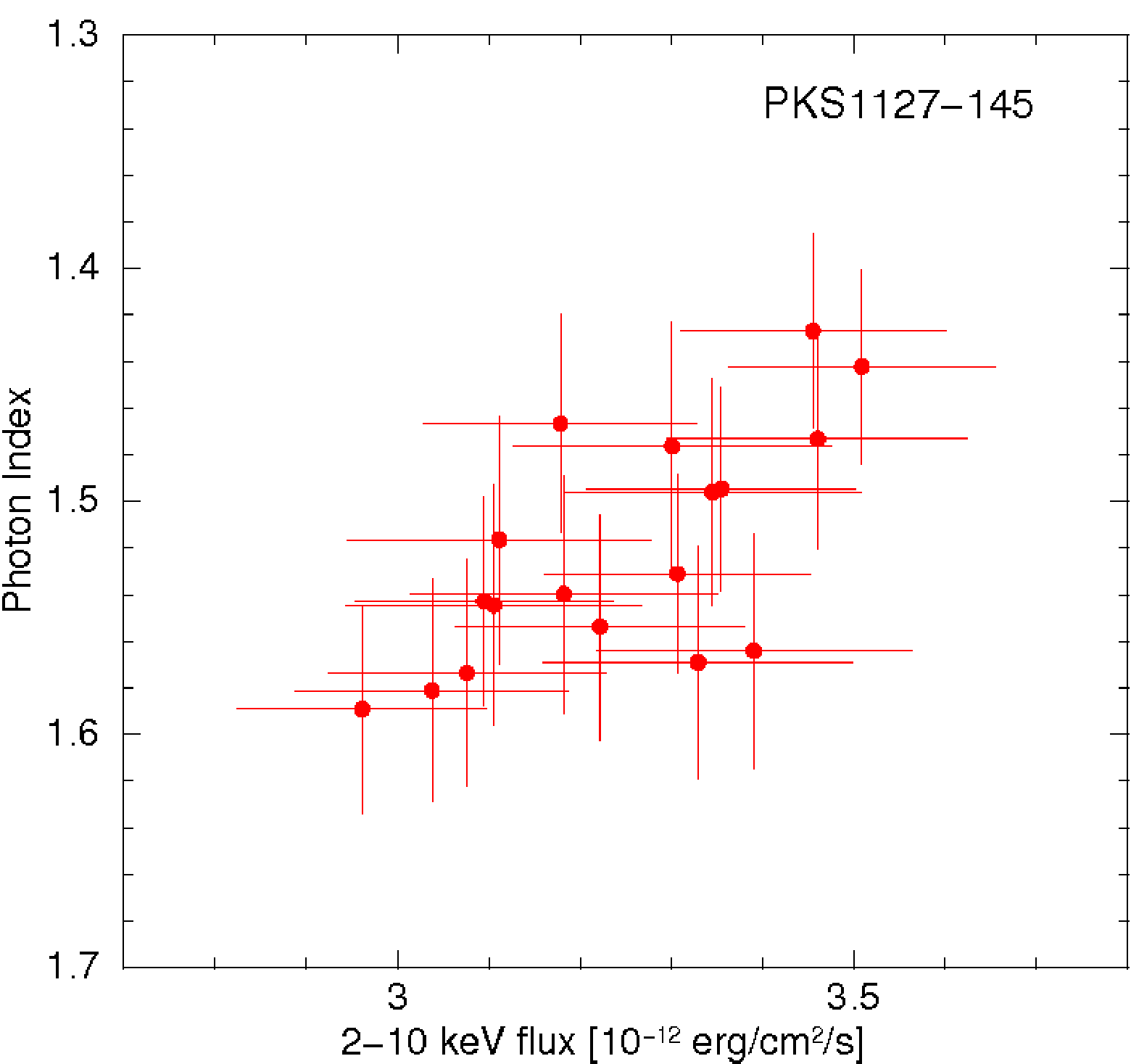}
\includegraphics[angle=0,scale=.3]{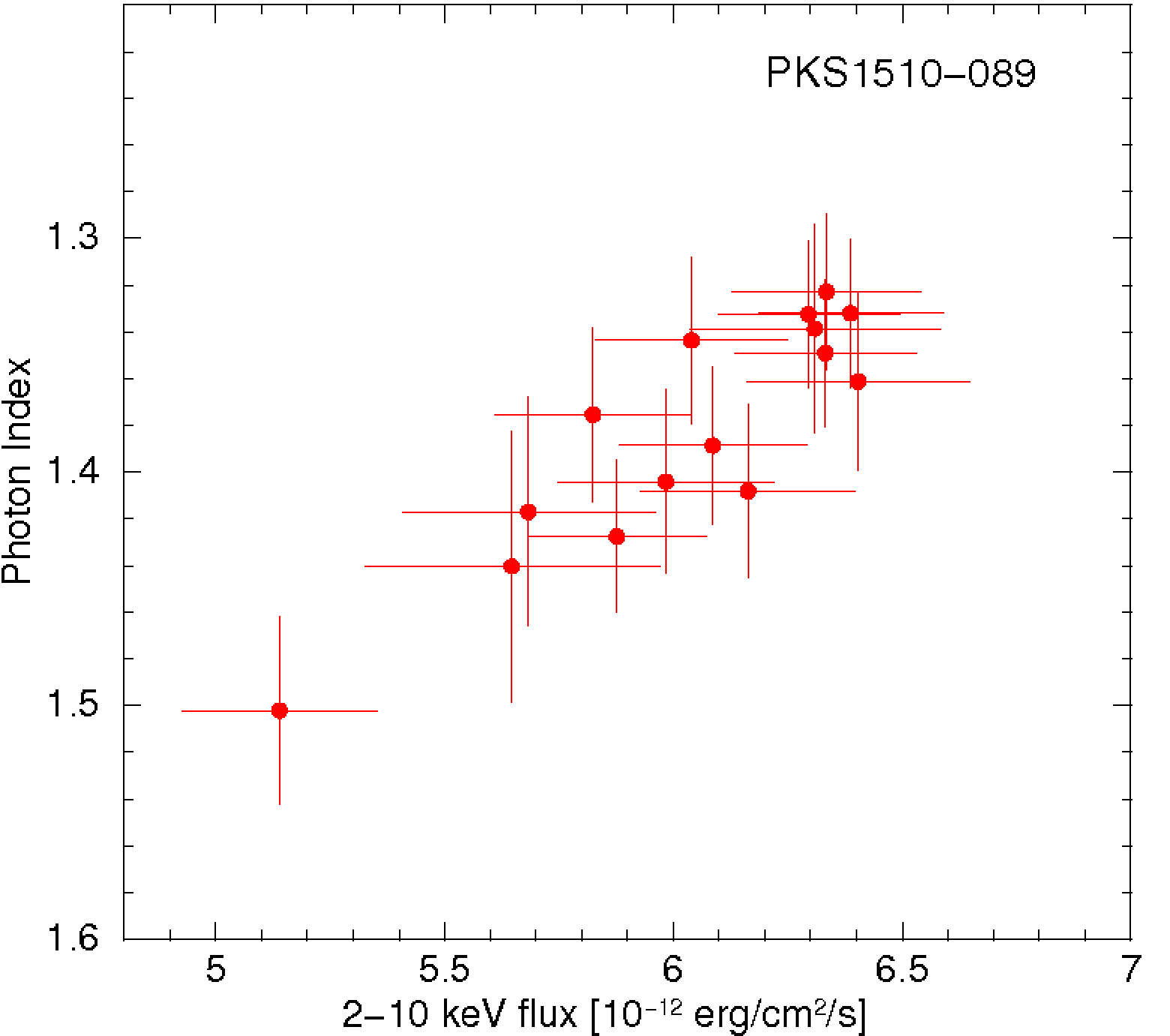}
\includegraphics[angle=0,scale=.3]{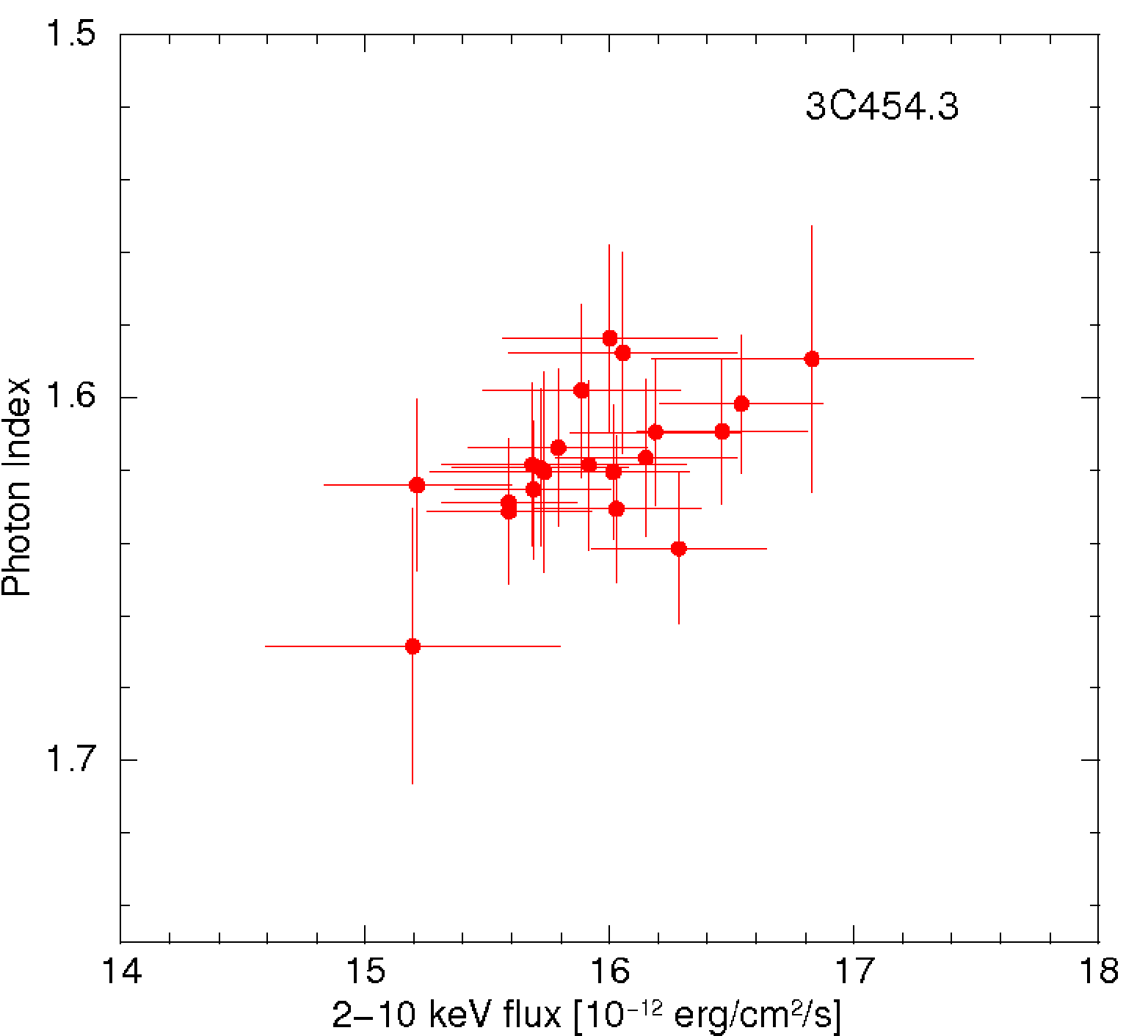}
\caption{Correlation of the $2-10$\,keV flux vs. photon index of five blazars 
as measured by the {\it Suzaku} XISs. \label{fig:fluxindex}}
\end{center}
\end{figure}

In order to investigate the X-ray spectral evolution of each object, 
we divided the total exposure into one-orbit intervals ($\sim 5760$\,s).
We fitted the overall XIS spectra between $0.3$ and $10$\,keV with an 
absorbed simple power-law function. The photoelectric absorbing 
column densities were fixed at the values derived in $\S$\,3.1.2.
Figure\,\ref{fig:fluxindex} shows the relation between the $2-10$\,keV 
fluxes versus the photon indices measured by the {\it Suzaku} XISs. 
Significant spectral variation is seen in PKS\,0208$-$512 ($\Gamma =1.4-1.8$), 
Q\,0827+243 ($\Gamma =1.2-1.6$), PKS\,1127$-$145 ($\Gamma =1.4-1.6$), and 
PKS\,1510$-$089 ($\Gamma=1.3-1.5$). In the case of 3C\,454.3, the X-ray
photon index is only weakly variable around the mean value $\Gamma \sim 1.6$.
Figure\,\ref{fig:fluxindex} clearly reveals a spectral evolution with 
the X-ray spectra hardening as the sources become brighter. Such a trend 
is often observed in high-frequency-peaked BL Lac objects (e.g., Kataoka et al. 1999), 
but it has never been observed so clearly in FSRQs (but see Kataoka et al. 2008 for PKS\,1510$-$089). 

\subsection{{\it Swift} }

Since the effective area of the $Swift$ XRT is less than 10\% of the $Suzaku$ XIS 
in the 0.5$-$10 keV range, detailed spectral modeling is difficult using $Swift$ data.
Furthermore, the average exposure for the $Swift$ observation was only a few 
kiloseconds, which was much less than the $Suzaku$ exposure.
We therefore fit the XRT data simply with a power-law model with Galactic absorption 
in the energy range 0.3$-$10 keV for the cross-calibration between the two instruments.
The results of the spectral fits are summarized in Table\,\ref{table:xrtfitresult}.
We can see that the results obtained with $Suzaku$ and $Swift$ are consistent 
within the range of error except PKS 1127$-$145.

\begin{table}[ht]
\footnotesize
\caption{Results of the spectral fits to the {\it Swift} XRT data using 
a power-law with Galactic absorption}
\begin{center}
\label{table:xrtfitresult}
\begin{tabular}{lccclc}
\tableline
Object      & $N_{\rm H}$\tablenotemark{a}  & $\Gamma$ & $F_{2-10 \rm keV}$\tablenotemark{b} & $\chi^2_{\rm r}$ \\
            &                               &          &                                     &    (dof)        \\
\tableline
0208$-$512 & 3.08 (fixed) & 1.96$\pm$0.24 & 1.37$\pm$0.06 & 0.33 (13) \\
0827+243 & 3.62 (fixed) & 1.46$\pm$0.35 & 1.13$_{-0.43}^{+0.55}$ & 1.12 (8) \\
1127$-$145 & 3.83 (fixed) & 1.28$\pm$0.03 & 6.14$\pm$0.23 & 1.28 (94) \\
1510$-$089 & 7.88 (fixed) & 1.38$\pm$0.08 & 6.09$_{-0.62}^{+0.65}$ & 0.71 (31) \\
3C\,454.3 & 7.24 (fixed) & 1.53$\pm$0.09 & 17.6$\pm$2.1 & 1.41 (18) \\
\tableline
\end{tabular}
\tablenotetext{}{Errors correspond to 1$\sigma$ confidence level.}
\tablenotetext{a}{Fixed value indicates the Galactic absorption column density in units of $10^{20}$\,cm$^{-2}$.}
\tablenotetext{b}{Flux in units of $10^{-12}$\,erg\,cm$^{-2}$\,s$^{-1}$.}
\end{center}
\end{table}

The UVOT fluxes in each filter were corrected for Galactic extinction following 
the procedure described in Cardelli et al. (1989). 
We generated a list of the amount of extinction that needs to be accounted 
for in each filter, $A_{\lambda}=E_{B-V}(aR_V + b)$, where $a$ and $b$ are 
constants. 
The Cardelli procedure provides a good approximation to the UV-through-IR Galactic dust 
extinction as a function of the total-to-selective extinction, $R_V$, 
which throughout this paper we assume to be $R_V=3.1$, which is the mean Galactic value.
The observed magnitudes and correction factors for each of 
the filters are summarized in Table\,\ref{table:Swiftmag} and Table\,\ref{table:correction}, respectively.
\begin{table}[ht]
\footnotesize
\caption{{\it Swift} UVOT magnitudes of five FSRQs}
\begin{center}
\label{table:Swiftmag}
\begin{tabular}{lccccccc}
\tableline
Object      & $v$   & $b$   & $u$   &  uvw1 &  uvm2 & uvw2  & E(B-V)  \\
            & (mag) & (mag) & (mag) & (mag) & (mag) & (mag) &  \\
\tableline
0208$-$512 & 17.64$^{+0.14}_{-0.12}$ & 17.94$^{+0.09}_{-0.08}$ & 17.17$^{+0.09}_{-0.08}$ & 16.84$\pm$0.07 & 16.71$\pm$0.07 & 17.00$^{+0.06}_{-0.05}$ & 0.022 \\
0827+243 & - & - & 16.57$\pm$0.01 & - & - & - & 0.033 \\
1127$-$145 & 16.48$\pm$0.02 & 16.70$\pm$0.01 & 15.64$\pm$0.01 & 15.51$\pm$0.01 & 15.56$\pm$0.01 & 15.79$\pm$0.01 & 0.037 \\
1510$-$089 & - & - & 16.01$\pm$0.02 & 16.27$\pm$0.02 & 16.13$\pm$0.02 & - & 0.097 \\
3C\,454.3  & 16.05$^{+0.09}_{-0.08}$ & 16.52$\pm$0.06 & 15.72$\pm$0.06 & 15.75$\pm$0.06 & 15.81$\pm$0.08 & 16.06$\pm$0.05 & 0.107 \\
\tableline
\end{tabular}
\tablenotetext{}{Observed magnitude for each observation using specific filter (Galactic extinction not corrected).}
\end{center}
\end{table}

\begin{table}[ht]
\footnotesize
\caption{Correction factors for the Galactic extinction in UV and optical filters}
\begin{center}
\label{table:correction}
\begin{tabular}{ll|ccccccc}
\tableline
Param &                                   & $v$ & $b$ & $u$ & uvw1 & uvm2 & uvw2 \\        
\tableline
$\lambda$\tablenotemark{a} & (nm)         & 547 & 439 & 346 & 260  & 249  & 193 \\
$a$\tablenotemark{b}       &              & 1.0015 & 0.9994 & 0.9226 & 0.4346 & 0.3494 & $-$0.0581 \\ 
$b$\tablenotemark{b}       &              & 0.0126 & 1.0171 & 2.1019 & 5.3286 & 6.1427 & 8.4402 \\
\tableline
                               & 0208$-$512 & 0.07 & 0.09 & 0.11 & 0.15 & 0.16 & 0.18 \\
                               & 0827+243 & -    & -    & 0.16 & -    & -    & -    \\ 
$A_{\lambda}$\tablenotemark{b} & 1127$-$145 & 0.12 & 0.15 & 0.18 & 0.25 & 0.27 & 0.31 \\
                               & 1510$-$089 & -    & -    & 0.48 & 0.65 & 0.70 & -    \\
                               & 3C\,454.3 & 0.33 & 0.44& 0.53 & 0.71 & 0.77 & 0.88 \\
\tableline
\end{tabular}
\tablenotetext{a}{Center wavelength for each optical and UV filter.}
\tablenotetext{b}{Parameters for calculating Galactic extinction for optical
and UV filters, calculated according to the prescription in Cardelli et al. (1989).
The Galactic reddening was taken from Schlegel et al. (1998).}
\end{center}
\end{table}

\subsection{{\it Fermi} LAT}

To study the average spectra of five objects during the four or five months 
of observations, we use the standard maximum-likelihood spectral estimator 
provided with the LAT science tools \textsc{gtlike}. This fits the data to 
a source model, along with models for the uniform extragalactic and structured 
Galactic backgrounds. Photons were extracted from a region with a 
$10^{\circ}$ radius centered on the coordinates of the position of 
each object. 
The Galactic diffuse background model is the currently recommended 
version (gll$\_$iem$\_$v02
\footnote{This model is available for download from the Fermi Science Support 
Center, http://fermi.gsfc.nasa.gov/ssc.}), 
with the normalization free to vary in the fit. 
The response function used is P6\_V3\_DIFFUSE.

For simplicity, we model the continuum emission from each source with a single power-law. 
It is likely that such a model might be too simple, as shown in the paper reporting 
spectra of bright Fermi blazars (Abdo et al. 2010),
where the gamma-ray data suggest a steepening of the spectrum with
energy, well-described as a broken power-law. However, here, we are reporting 
cases of blazars in low-level activity states and thus relatively faint, 
where fits to a broken power-law model would result in poorly constrained 
spectral parameters for a more complex model;  
furthermore, we note that the use of such more
complex spectral model in the gamma-ray band does not alter our conclusions or
significantly change the parameters in Table 10.  
The extragalactic background is assumed to have a power-law spectrum, 
with its spectral index and the normalization free to vary in the fit. 
From an unbinned \textsc{gtlike} fit the best fit photon indices 
are $\Gamma = 2.33\pm0.05$ for PKS\,0208$-$512, 
$\Gamma = 2.62\pm0.35$ for Q\,0827+243, 
$\Gamma = 2.77\pm0.14$ for PKS\,1127$-$145, 
$\Gamma = 2.48\pm0.03$ for PKS\,1510$-$089, and 
$\Gamma = 2.51\pm0.02$ for 3C\,454.3 (see also Table\,\ref{table:latresult}). 
Here only statistical errors are taken into account, and we report fluxes 
using spectra extrapolated down to $100$\,MeV. In the case of bright sources 
(PKS\,1510$-$089 and 3C\,454.3), we also analyzed the data collected during 
the {\it Suzaku} observing period to construct the simultaneous 
broad-band spectra spectra. 

Table \ref{table:latfluxresult} summarizes the flux in seven energy bands 
obtained by separately running \textsc{gtlike} for each energy band;
200$-$400 keV, 400$-$800 keV, 800$-$1600 keV, 1600$-$3200 keV, 3200$-$6400 keV, 
6400$-$12800 keV, 12800$-$25600 keV, respectively.

\begin{table}[ht]
\footnotesize
\caption{Results of the spectral fits to the {\it Fermi} LAT data}
\begin{center}
\label{table:latresult}
\begin{tabular}{lccc}
\tableline
Object      & $\Gamma$ & $F_{> 100 \rm MeV}$\tablenotemark{a}   & TS\tablenotemark{b}\\
            &          &                       &       \\
\tableline
0208$-$512 & 2.33$\pm$0.05  & 0.26$\pm$0.03 & 1484 \\
0827+243 & 2.62$\pm$0.36  & 0.05$\pm$0.04 & 58 \\
1127$-$145 & 2.75$\pm$0.14  & 0.15$\pm$0.04 & 234 \\
1510$-$089 & 2.48$\pm$0.03  & 0.69$\pm$0.04 & 4224 \\
3C\,454.3 & 2.50$\pm$0.02 & 2.55$\pm$0.08 & 25144 \\
\tableline
1510$-$089\tablenotemark{c}  & 2.28$\pm$0.27 & 0.91$\pm$0.51 & 59 \\
3C\,454.3\tablenotemark{c} & 2.62$\pm$0.13 & 2.59$\pm$0.58 & 281\\
\tableline
\end{tabular}
\tablenotetext{a}{Flux in units of 10$^{-6}$\,ph\,cm$^{-2}$\,s$^{-1}$.}
\tablenotetext{b}{Test statistic: defined as TS\,$=2(\log L - \log L_0)$, 
where $L$ and $L_0$ are the likelihood when the source is included or not.}
\tablenotetext{c}{Corresponding data collected during the {\it Suzaku} observing period.}
\end{center}
\end{table}

\begin{table}[ht]
\footnotesize
\caption{Results of {\it Fermi} LAT data analysis from 200 MeV to 25600 MeV
(Flux in units of 10$^{-9}$\,ph\,MeV$^{-1}$\,cm$^{-2}$\,s$^{-1}$)}
\begin{center}
\label{table:latfluxresult}
\begin{tabular}{lccccccc}
\tableline
Object      & Band 1\tablenotemark{a} & Band 2\tablenotemark{a} & Band 3\tablenotemark{a} & Band 4\tablenotemark{a} 
& Band 5\tablenotemark{a} & Band 6\tablenotemark{a} & Band 7\tablenotemark{a}\\
\tableline
0208$-$512 & 3.24$\pm$0.25 & 3.69$\pm$0.30 & 4.19$\pm$0.42 & 3.81$\pm$0.61 & 2.88$\pm$0.79 & 1.95$\pm$1.00 & 1.51$\pm$1.49 \\
0827+243 & 0.73$\pm$0.22 & 1.03$\pm$0.28 & 1.08$\pm$0.36 & 0.92$\pm$0.50 & - & - & - \\
1127$-$145 & 2.65$\pm$0.34 & 2.50$\pm$0.43 & 3.21$\pm$0.68 & 3.17$\pm$1.05 & - & - & - \\
1510$-$089 & 9.74$\pm$0.38 & 10.62$\pm$0.48 & 10.19$\pm$0.67 & 11.57$\pm$1.09 & 7.51$\pm$1.47 & 7.15$\pm$2.26 & 5.14$\pm$3.01 \\
3C\,454.3 &37.22$\pm$0.67 & 39.36$\pm$0.91 & 43.79$\pm$1.42 & 44.20$\pm$2.31 & 26.65$\pm$2.96 & 17.98$\pm$3.99 & 4.55$\pm$3.28 \\
\tableline
\end{tabular}
\tablenotetext{a}{Band 1: 200$-$400 MeV, Band 2: 400$-$800 MeV, Band 3: 800$-$1600 MeV, Band 4: 1600$-$3200 MeV, 
Band 5: 3200$-$6400 MeV, Band 6: 6400$-$12800 MeV, and Band 7: 12800$-$25600 MeV}\\
\end{center}
\end{table}

\section{Discussion}

\subsection{Broad-Band Spectra spectral fits}

\begin{figure}
\begin{center}
\includegraphics[scale=.5]{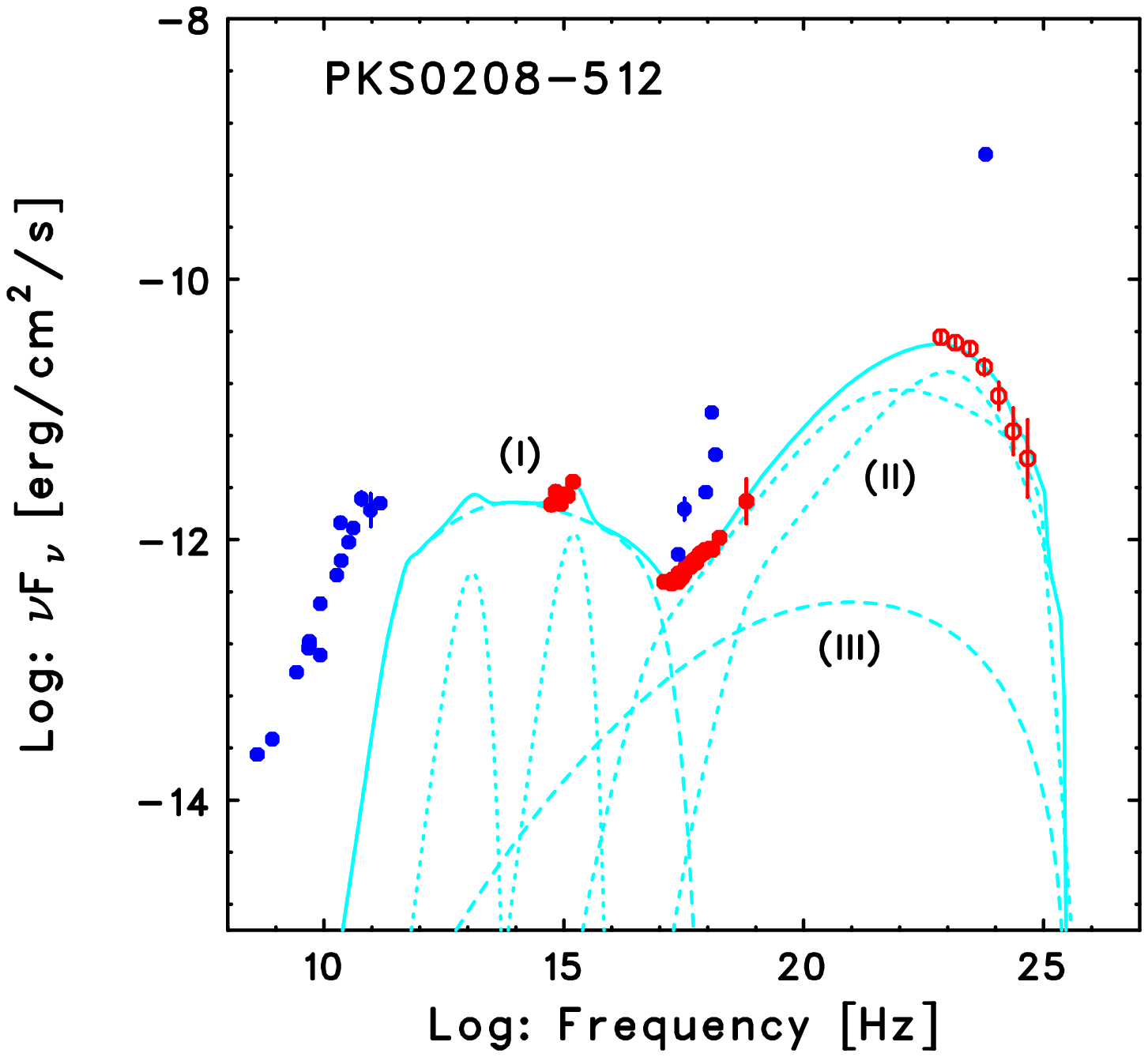}
\includegraphics[scale=.5]{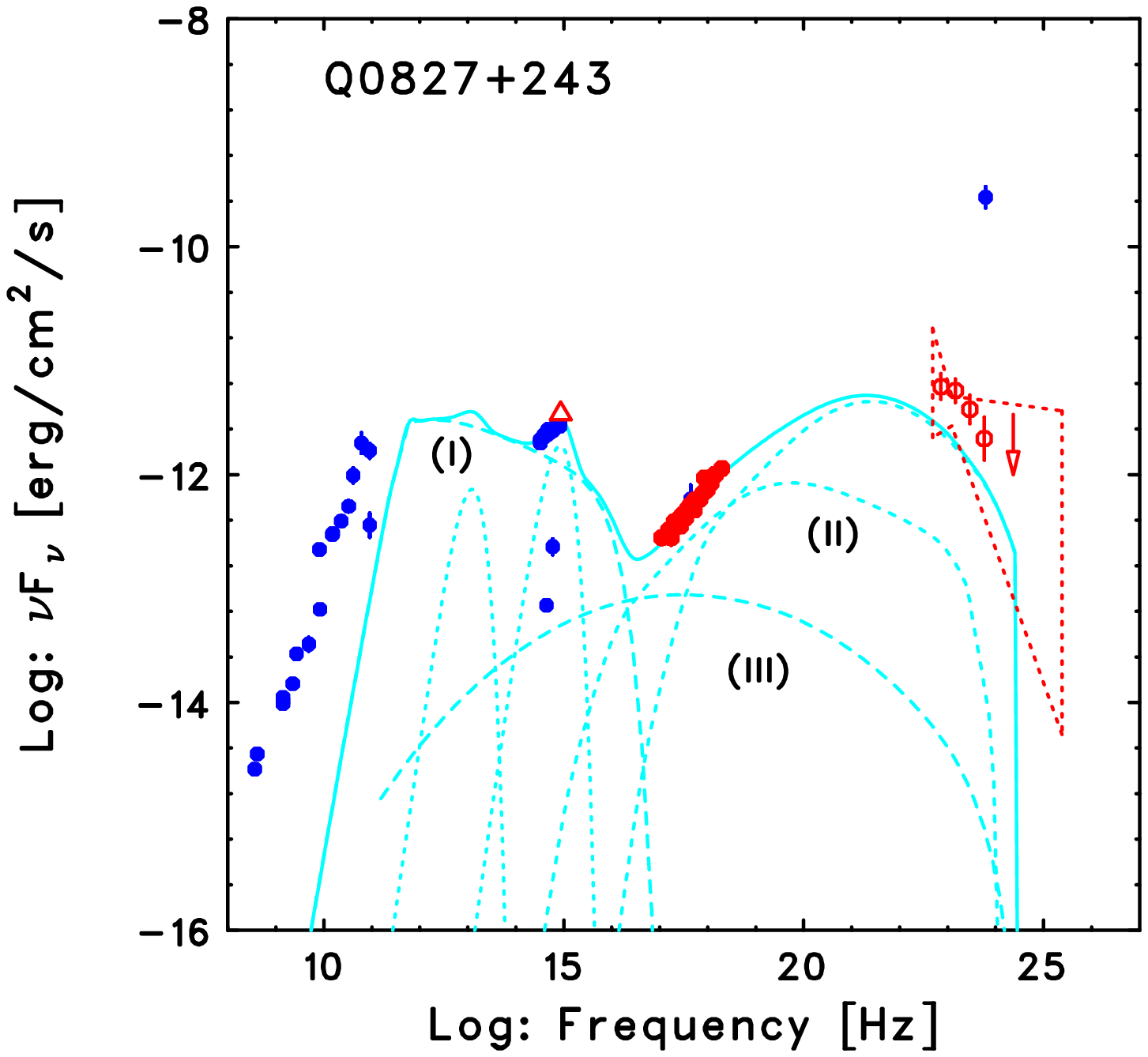}
\includegraphics[scale=.5]{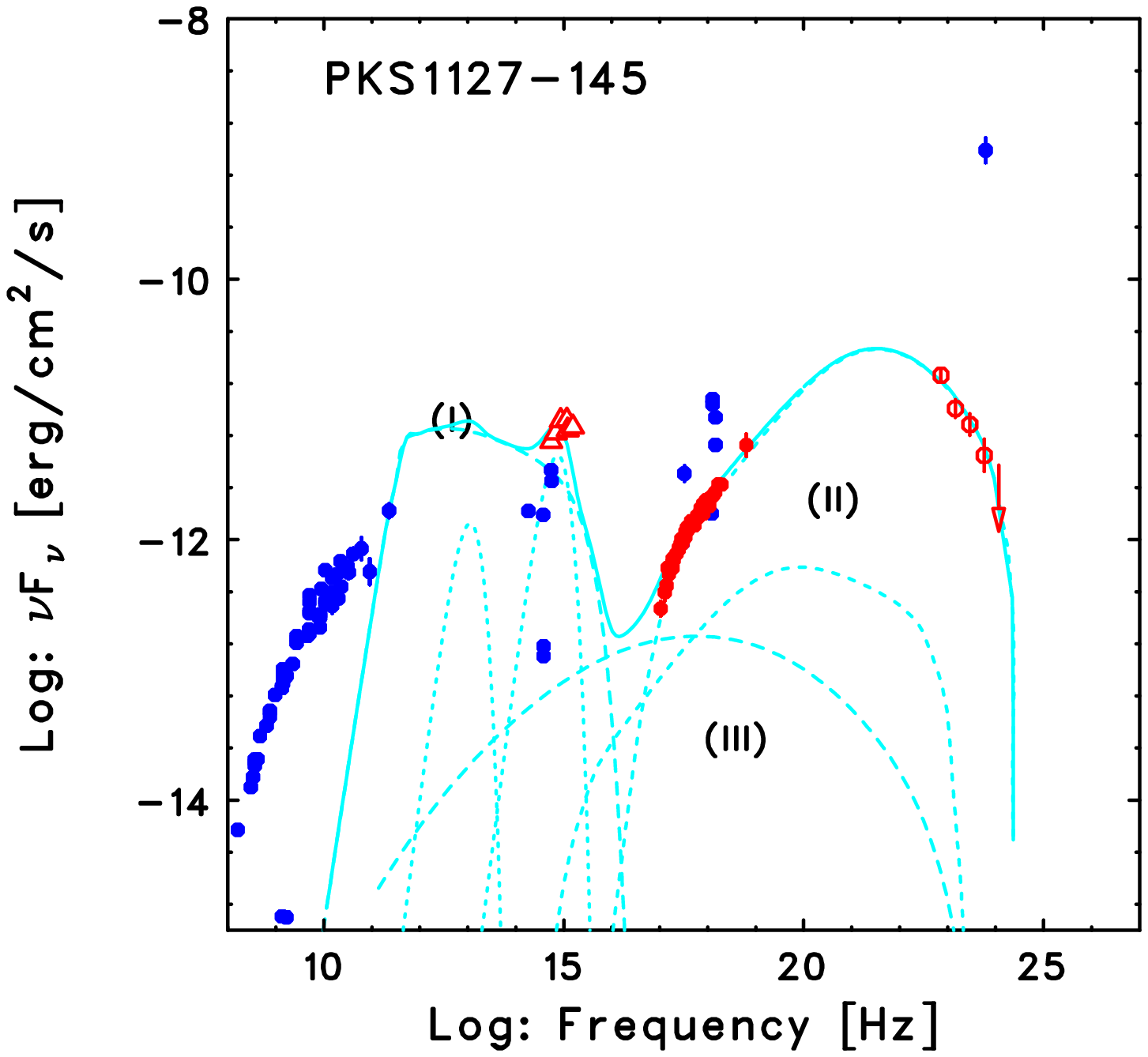}
\includegraphics[scale=.5]{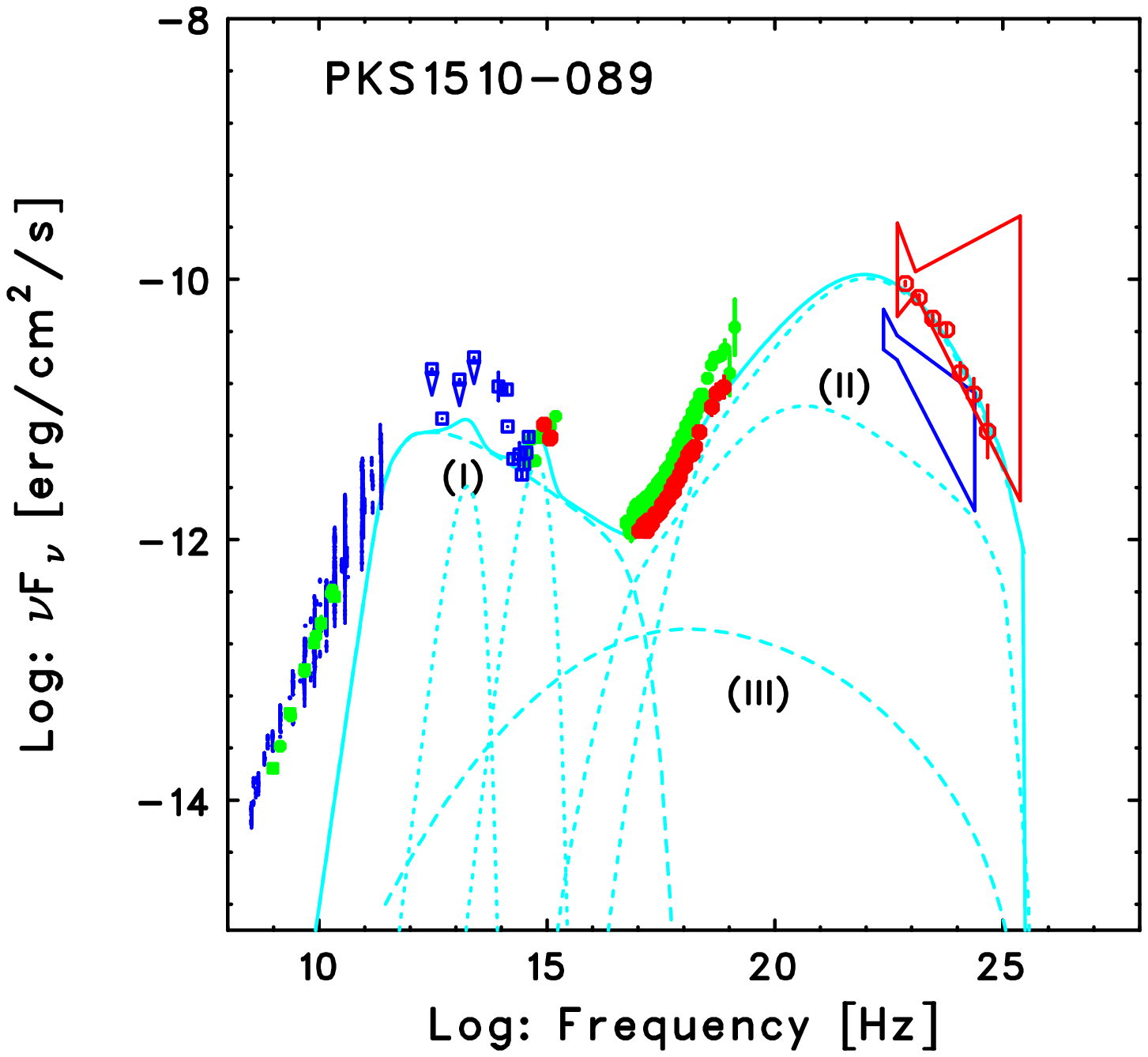}
\includegraphics[scale=.5]{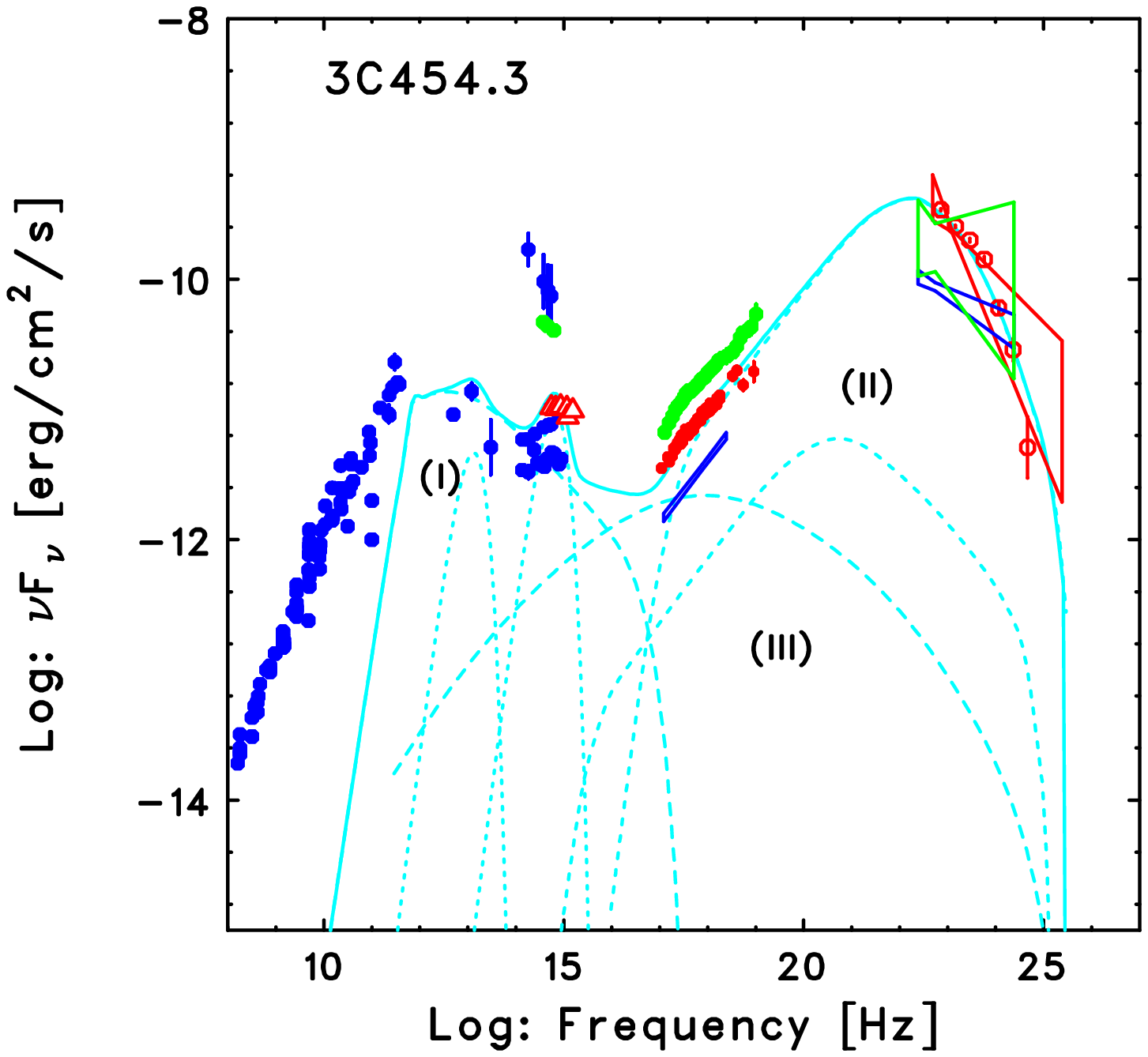}
\caption{Overall SED of five sources constructed with broad-band data obtained
during 2008 October to 2009 January (filled red circles and solid bow-tie). 
Quasi simultaneous data are also shown (open red circles).
Historical radio (NED) and $\gamma$-ray (EGRET) data
are also plotted as filled blue circles. Green symbols in the SEDs of
PKS\,1510$-$089 and 3C\,454.3 denote the previous simultaneous observations
(Kataoka et al. 2008; Donnarumma et al. 2010). 
The dotted lines show (I) the synchrotron and (II) the EC components, 
and (III) SSC components, respectively. 
The solid line shows the jet continuum calculated
with the jet emission model described in $\S$\,4.2.}
\label{fig:nufnu}
\end{center}
\end{figure}

We constructed the broad-band spectral energy distribution (SED) 
ranging from the radio to $\gamma$-ray bands for the five observed FSRQs, 
and these are shown in Figure\,\ref{fig:nufnu}. Here the filled red circles and solid lines 
represent simultaneous data from the UV/optical ({\it Swift} UVOT), 
X-ray ({\it Suzaku}) and $\gamma$-ray ({\it Fermi} LAT) observations.
Quasi-simultaneous data are also shown as red open triangles and 
dashed lines. Historical radio (NED) and $\gamma$-ray (EGRET) data
are also plotted as filled blue circles. Green symbols in the SEDs of 
PKS\,1510$-$089 and 3C\,454.3 denote the previous simultaneous observations 
(Kataoka et al. 2008; Donnarumma et al. 2010). 

In order to model the constructed SEDs, we applied the synchrotron-inverse 
Compton (IC) emission model described in Tavecchio \& Ghisellini (2008), 
where both synchrotron and external (BLR and DT) photons are considered as 
seed radiation fields contributing to the IC process (SSC+ECR).
The electron distribution is modeled as a smoothly broken power-law:
\begin{equation}
N'(\gamma) = K \, \gamma^{-n_1} \left(1 + \frac{\gamma}{\gamma_{\rm br}} \right)^{n_1 - n_2} \, ,
\end{equation}
\noindent
where $K$ (cm$^{-3}$) is a normalization factor, $n_1$ and $n_2$ are the energy
indices below and above the break Lorentz factor $\gamma_{\rm br}$.
The electron distribution extends within the limits
$\gamma_{\rm min} < \gamma < \gamma_{\rm max}$. We also assume that the `blazar 
emission zone', with the comoving size $R$ and magnetic field intensity $B$, is 
located at the distance $r$ such that $r_0 < r < r_{\rm BLR} < r_{\rm DT}$, where $r_0$ is 
the distance below which the photon energy density in the jet rest frame is 
dominated by the direct radiation of the accretion disk, $r_{\rm BLR}$ is 
the characteristic scale of the broad-line-region, and $r_{\rm DT}$ is the
scale of the dusty torus (see the discussion in Tavecchio \& Ghisellini 2008 
as well as in Sikora et al. 2009). This choice, while somewhat arbitrary, 
has been validated by a number of authors modeling broad-band spectra of FSRQs.  
Hence, the comoving energy density of the dominant photon
field --- provided by the BLR --- is
\begin{equation}
U'_{\rm rad} \simeq \Gamma_j^2 \, {\eta_{\rm BLR} L_{\rm d} \over 4 \pi r_{\rm BLR}^2 c} \, ,
\end{equation}
\noindent
where $\Gamma_j$ is the jet bulk Lorentz factor, and the BLR is assumed to reprocess 
$\eta_{\rm BLR} \simeq 10\%$ of the disk luminosity $L_{\rm d}$. 
Finally, we assume that the jet viewing angle is in all the cases $\theta_j \simeq 1 / \Gamma_j$,
so that the jet Doppler factor $\delta_j \simeq \Gamma_j$. 

The results of model fitting are shown in different panels of Figure\,\ref{fig:nufnu}, and the 
resulting parameters are summarized in Table\,\ref{table:SEDresult}. 
In the context of this model, where we assume the dissipation region 
to be between the immediate vicinity of
the accretion disk but within the BLR, it is clear that in all cases
the LAT fluxes are dominated by the IC/BLR component, while in the X-ray band both IC/BLR and
IC/DT processes may contribute at a comparable level. In addition, the SSC emission seems negligible,
being in particular too weak to account for the soft X-ray excess discussed in the previous 
sections. This excess, on the other hand, may be well represented by the high-energy tail of
the synchrotron continuum, or an additional blackbody-type spectral component. 

\begin{table}[ht]
\footnotesize
\caption{Model parameters used to calculate the SEDs of five FSRQs}
\begin{center}
\label{table:SEDresult}
\begin{tabular}{lccccccccccccc}
\tableline
Object   & $n_1$ & $n_2$ & $\gamma_{\rm min}$ & $\gamma_{\rm br}$ & $\gamma_{\rm max}$ & $K$                 & $\Gamma_j$ & $R$            & $B$ & $L_{\rm d}$       & $r_{\rm BLR}$ & $r_{\rm DT}$\\
         & & &            &                   &   &  [$10^4$\,cm$^{-3}$] &          & [$10^{16}$\,cm] & [G] & [$10^{46}$\,erg/s] & [$10^{18}$\,cm]& [$10^{18}$\,cm]\\
\tableline
0208$-$512 & 2 & 3.3 & 3.0 & 700 & 4.3\,$\times 10^4$ & 2.2  & 15 & 1.8 & 1.1  & 1.5 & 0.76 & 3.0\\
0827+243 & 2 & 3.3 & 1.5 & 300 & 1.0\,$\times 10^4$ & 8.5  & 10 & 1.8 & 3.8  & 2.0 & 1.3  & 4.2\\
1127$-$145 & 2 & 3.4 & 1.2 & 110 & 5.0\,$\times 10^3$ & 0.65 & 10 & 6.5 & 4.1  & 10  & 0.8  & 10\\
1510$-$089 & 2 & 3.5 & 3.0 & 190 & 4.6\,$\times 10^4$ & 0.81 & 13 & 4   & 0.8  & 0.3 & 0.48 & 4\\
3C\,454.3 & 2 & 3.8 & 1.0 & 290 & 3.0\,$\times 10^4$ & 4.5 & 12 & 3.2 & 0.8  & 4.0 & 1.5  & 30\\
\tableline
\end{tabular}
\end{center}
\end{table}

Based on the model results, for each object we compute the ratio of 
the comoving energy densities stored in jet electrons and the magnetic field, 
\begin{equation}
{U'_e \over U'_B} = {\int_{\gamma_{\rm min}}^{\gamma_{\rm max}} \gamma \, m_e c^2 \, 
N'(\gamma) \, d\gamma \over B^2 / 8 \pi} \, ,
\end{equation}
\noindent
where $B$ is the magnetic field intensity in the emission region. 
In addition, we compute the implied total kinetic jet power as
\begin{equation}
L_j = \pi R^2 c \Gamma_j^2 \, \left(U'_e + U'_B + U'_p\right) \, ,
\end{equation}
\noindent
where $R$ is the emission region linear size, and $U'_p$ is the energy density of cold
protons. The latter parameter is estimated assuming one proton per ten electron-positron 
pairs (see the discussion in Sikora et al. 2009), namely $U'_p = 0.1 \, m_p c^2 
\int_{\gamma_{\rm min}}^{\gamma_{\rm max}} N'(\gamma) \, d\gamma$. The resulting
total kinetic power of the outflow is then compared with the accretion luminosity
(assuming standard accretion disk with $10\%$ radiative efficiency), by means of the evaluated
efficiency parameter $\eta_j = L_j / L_{\rm acc} \simeq L_j / 10 \, L_{\rm d}$, where $L_{\rm d}$ 
is the disk luminosity implied by the model fitting (see Table\,\ref{table:SEDresult}).
Note that with the above model assumptions and the model parameters inferred
by us, the jets of objects considered here 
are dynamically dominated by cold protons, $U'_p / U'_e \simeq 200 / \langle \gamma \rangle > 1$, 
since the mean Lorentz factor of the radiating ultra-relativistic electrons is in all the cases
$\langle \gamma \rangle \ll 200$ (see Table\,\ref{table:jet}).

\begin{table}[ht]
\footnotesize
\caption{Jet parameters of five FSRQs in low-activity states}
\begin{center}
\label{table:jet}
\begin{tabular}{lcccc}
\tableline
Object   & $U'_e/U'_B$ & $L_j$ & $\eta_j$ & $\langle \gamma \rangle$\\
         &             & [$10^{46}$\,erg/s] & & \\
\tableline
0208$-$512 & 2    & 0.8 & 0.06 & 16\\
0827+243 & 0.6  & 2.8 & 0.14 & 8\\
1127$-$145 & 0.03 & 5.8 & 0.06 & 5\\
1510$-$089 & 0.9  & 1   & 0.35 & 12\\
3C\,454.3 & 7   & 9.4 & 0.23 & 5\\
\tableline
\end{tabular}
\end{center}
\end{table}

Some of the derived jet parameters for five luminous blazars in their
low-activity states are significantly different from the
analogous parameters claimed for the flaring states, {\sl even in the same object}.
For example, in the case of the high-activity state of PKS\,1510$-$089,
Kataoka et al. (2008) estimated (under the same assumptions regarding the jet content 
as in this paper) the total kinetic power of the jet as 
$L_j \sim 2.7 \times 10^{46}$\,erg\,s$^{-1}$,
which is larger than the value derived in this paper, by about a factor of 3.  
In addition, our model values of the jet bulk Lorentz factors are also systematically
lower than the ones given in the literature ($\Gamma_j \simeq 10$ versus $20$).
Interestingly, other jet parameters, such as magnetic field intensity - $B \simeq 1$\,G - 
and the equipartition ratio, $U'_e/U'_B \sim 1$, 
or the general spectral shape of the electron energy distribution, 
are comparable to the ones found for flaring FSRQs, (albeit with a substantial scatter).    
It should be noted in this context, however, that for the three sources
considered in this paper (namely PKS\,1127$-$145, PKS\,1510$-$089, and 3C\,454.3), the
flaring states were analyzed in a framework of the IC/DT model (B\l a\.zejowski et al.
2004, Kataoka et al. 2008, Sikora et al. 2008, respectively), 
while here, we argue that the IC/BLR contribution is dominant, 
as motivated by the detected relatively short (day) variability timescale of the X-ray continua.
On the other hand, as discussed recently in
Sikora et al. (2009), there is so-called a `conspiracy' between the IC/BLR and IC/DT
models, in a sense that the resulting inferred jet parameters are comparable in both cases.
Hence, we can safely conclude that the low- and high-activity states of
luminous blazar sources are due to the low and high total kinetic power of the jet, respectively, 
possibly related to varying bulk Lorentz factors within the blazar
emission zone. And indeed, keeping in mind that the highly 
dynamical and complex jet formation
processes in the closest vicinity of supermassive black holes -- most likely 
shaped by accretion process subjected to several possible instability 
of the jet fuel, especially when 
the accretion rate is close to Eddington -- such a significant 
variation in the total kinetic output of the outflow should not be surprising.
Further support for this scenario comes from the fact that the jet efficiency factors
estimated here, $\eta_j \lesssim 1$, are significantly lower than the
ones found for powerful blazars in their flaring states (see Sambruna et al. 2006, 
Ghisellini et al. 2009), even if the difference in the jet proton content 
adopted by various authors is taken into account.  

\subsection{Spectral Evolution}

As shown in $\S$\,3.1.3, the X-ray spectra of the FSRQs analyzed here flatten 
with increasing flux. 
For $\Gamma_{\rm j}\sim\delta_{\rm j}\sim10$ and the dominant IC/BLR emission 
process, the electrons emitting the observed 1$-$10 keV photons have 
Lorentz factor $\gamma\sim\gamma_{\rm min}\sim$ few.
The electrons emitting X-ray photons in these sources 
are very low-energy, so cooling effects cannot play any role in the
observed spectral evolution. In particular, it can be easily demonstrated that
in a framework of our model (i.e., for the dominant IC/BLR energy losses), 
a strong cooling regime is expected only for the electrons with Lorentz factors 
greater than
\begin{eqnarray}
\gamma_{\rm cr} & \simeq & {3 \pi \, m_e c^3 \, r_{\rm BLR}^2 \over \sigma_{\rm T} \, R \, \Gamma_j^2
\, \eta_{\rm BLR} L_{\rm d} } \nonumber \\
& \simeq & 350 \, \left({r_{\rm BLR} \over 10^{18}\,{\rm cm}}\right)^2
\left({\eta_{\rm BLR} \over 0.1}\right)^{-1} \left({\Gamma_j \over 10}\right)^{-2}
\left({L_{\rm d} \over 10^{46}\,{\rm erg/s}}\right)^{-1} \left({R \over 10^{16}\,{\rm cm}}\right)^{-1} \, .
\end{eqnarray}
\noindent
This, for the fitting parameters as given in Table\,\ref{table:SEDresult}, is 
typically above or just around the break Lorentz factor, $\gamma_{\rm cr} \gtrsim 
\gamma_{\rm br}$ (in agreement with the
discussion in Sikora et al. 2009). Adiabatic losses, if present, should not result 
in changing the slope of the power-law X-ray continua as well. Thus, one
may suspect that the revealed spectral changes are shaped by the acceleration 
process within the blazar emission zone. In the case of relativistic jets the 
relevant acceleration processes are still quite uncertain, although, as pointed 
out by Kataoka et al. (2008) and Sikora et al (2009), the repeatedly observed flat 
X-ray photon indices $\Gamma \leq 1.5$ seem to favor the mechanism discussed by 
Hoshino et al. (1992) for the low-energy segment of the electron energy distribution.
In this model, the low-energy electrons (with Lorentz factors, roughly, $\gamma < 
m_p/m_e$) are accelerated by a resonant absorption of the cyclotron emission
generated by cold protons reflected from the shock front. As shown later by
Amato \& Arons (2006), the power-law slope of these accelerated electrons depends 
on the relative number of electrons to protons at the shock front. Hence, a larger
fraction of the energy carried by jet protons during the higher-activity states
should in principle result in a more efficient acceleration of jet electrons and 
their flatter spectrum, in agreement with the observed X-ray spectral evolution 
discussed here.

The above interpretation, on the other hand, would imply a significant
variability in the $\gamma$-ray frequency range. Indeed, the broken
power-law form of the electron energy distribution revealed by our spectral
modeling discussed in the previous section implies the $\gamma$-ray flux 
$F_{\gamma} \equiv [\nu F_{\nu}]_{\gamma}$ around the IC spectral peak 
$\nu_{\gamma} \sim 10^{22}$\,Hz should be, roughly
\begin{equation}
F_{\gamma} \simeq F_{X} \, \left({\nu_{\gamma} \over \nu_X}\right)^{2-\Gamma} \! 
\simeq 10^{4\,(2-\Gamma)} F_{X} \, ,
\end{equation}
\noindent
where $F_{X}$ is the monochromatic X-ray flux measured around 
$\nu_X \sim 10^{18}$\,Hz, and $\Gamma$ is the observed X-ray photon index. 
For example, our analysis for PKS\,0208$-$512 indicates a photon index $\Gamma_1 \sim 1.8$ 
for an X-ray flux $F_{X,\,1 }\sim 1.2\times10^{-12}$\,erg\,cm$^{-2}$\,s$^{-1}$ 
in the lower state, and $\Gamma_2 \sim 1.5$ for $F_{X,\,2} \sim 1.6\times
10^{-12}$\,erg\,cm$^{-2}$\,s$^{-1}$ in the higher state. Thus, if the observed X-ray 
variability is due to flattening of the electron energy distribution during the 
acceleration process, one should observe the $\gamma$-ray variability of the
order of
\begin{equation}
{F_{\gamma,\,2} \over F_{\gamma,\,1}} \simeq 
10^{4\,(\Gamma_1 - \Gamma_2)} \, {F_{X,\,2} \over F_{X,\,1}} \sim 20 \, .
\end{equation}
\noindent
However, during the simultaneous {\it Fermi} observation, no significant $\gamma$-ray 
variability was observed for the analyzed sources, at least within one day timescale.

Therefore, the most viable explanation for the observed X-ray spectral evolution is 
that the IC power-law slope remains roughly constant during the flux variations, 
but the amount of contamination from the additional soft X-ray component increases
at low flux levels, affecting the spectral fitting parameters at higher photon energies 
($>2$\,keV). Note that in such a case the expected gamma-ray variability should be of 
the same order as the X-ray variability, namely $F_{\gamma,\,2}/F_{\gamma,\,1} \simeq 
F_{X,\,2}/F_{X,\,1} \sim 1.3$. 

We finally note in this context that, as shown in $\S$\,3.1.2, the previous BeppoSAX 
data for PKS\,0208$-$512 collected during the high state indicated a convex X-ray spectrum,
and an excess absorption below $1$\,keV with a column density of $N_{\rm H} \sim 1.67
\times10^{21}$\,cm$^{-2}$ exceeding the Galactic value by more than a factor of 5.
However, the X-ray photon index was similar to the one implied by our {\it Suzaku} 
observations ($\Gamma \sim 1.7$). Therefore, the convex spectrum observed by BeppoSAX 
may reflect an intrinsic shape of the IC emission involving the low-energy cut-off in the
electron energy distribution around $\gamma \sim 1$, as expected in the EC/BLR model 
(Tavecchio et al. 2007), which is only diluted during the low-activity states due to
the presence of an additional soft X-ray spectral component.

\begin{figure}[ht]
\begin{center}
\includegraphics[angle=0,scale=.5]{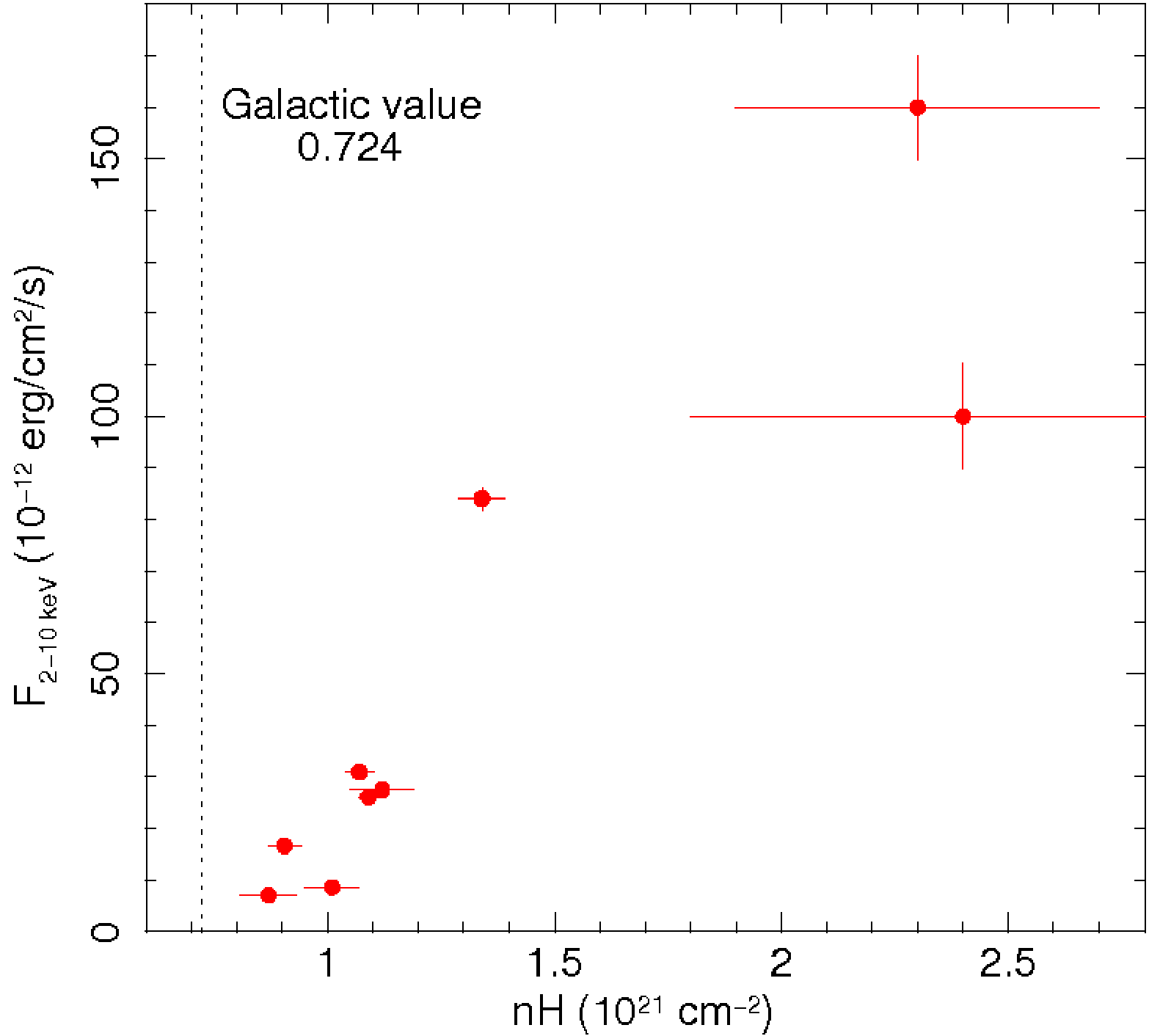}
\caption{Fluxes in the $2-10$\,keV band for different observations of 3C\,454.3 vs. $N_{\rm H}$ for the fits with an absorbed power-law model. 
The dashed line indicates the Galactic absorption column. 
This figure indicates that the intrinsic X-ray spectrum is not a simple power-law, 
but instead, it shows some curvature, which may depend on the X-ray brightness. \label{fig:nH-flux}}
\end{center}
\end{figure}

Similar trend has been observed in 3C\,454.3. To illustrate this, in Figure\,\ref{fig:nH-flux} 
we selected the data which have a similar power-law slope ($\Gamma \sim 1.6$) and plotted 
the absorption column versus $2-10$\,keV flux densities derived from the {\it Chandra} 
(Villata et al. 2006), {\it Swift} (Giommi et al. 2006), XMM-{\it Newton} (Raiteri et al. 2007, 2008), 
and {\it Suzaku} (this work) observations. We can see that there is a trend of increasing 
the absorption value with source brightness, as previously reported by Raiteri et al. (2007; 2008). 
These results may again be explained by the soft excess emission being more important 
when the source gets fainter, and becoming almost completely ``hidden" behind the hard X-ray 
power-law when the source gets brighter. 

From the spectral fitting of the {\it Suzaku} data, we showed in $\S$\,3.1.2 that the soft 
X-ray excess may be represented either by a steep power-law ($\Gamma\sim3-5$) or a 
blackbody-type emission ($kT\sim0.1-0.2$\,keV). Since the synchrotron peak of each 
source is located around optical photon energies (see Figure\,\ref{fig:nufnu}), 
the high-energy synchrotron tail may possibly account for the observed soft X-ray excess 
emission, especially if being modified by the Klein-Nishina effects (see the discussion
in Sikora et al. 2009 and Kataoka et al. 2008). On the other hand, the bulk-Compton 
spectral component produced by Comptonization of the UV accretion disk by cold electrons 
in the innermost parts of relativistic jets (e.g., Begelman \& Sikora 1987) is a natural
explanation for the apparent soft X-ray excess component. 

\section{Conclusions}

We have presented the observations and analysis of the data for the $\gamma$-ray-loud blazars, 
PKS\,0208$-$512, Q\,0827+243, PKS\,1127$-$145, PKS\,1510$-$089, and 3C\,454.3, obtained with the 
{\it Suzaku}, {\it Swift} UVOT and {\it Fermi} LAT. 
Observations were conducted between 2008 October and 2009 January. 
These observations allowed us to 
construct broadband spectra of the sources in the low $\gamma$-ray activity state, 
covering optical to GeV photon energy range. 
Our results are as follows: 

\begin{enumerate}

\item The X-ray spectra of five FSRQs are well 
represented by an absorbed hard power-law model ($\Gamma\sim1.4-1.7$). 
For PKS\,0208$-$512, PKS\,1127$-$145, and 3C\,454.3, the fitted absorption column
is larger than the Galactic value (but we note that the ``excess absorption'' is 
not a unique representation of X-ray spectra of those blazars).  
Compared with previous X-ray observations, we see a trend of increasing apparent 
X-ray absorption column with increasing high-energy luminosity of the source.  

\item {\it Suzaku} observations reveal spectral evolution of the X-ray emission: 
the X-ray spectrum becomes harder as the source gets brighter. 
Such spectral changes are most likely due to the underlying and steady low-energy 
spectral component which becomes prominent when the inverse-Compton emission gets 
fainter. This soft X-ray excess can be explained as a contribution of 
the high-energy tail of the synchrotron component, or bulk-Compton radiation. 

\item We adopt the location of the blazar emission region to be 
outside of the immediate vicinity of
the accretion disk but within the BLR, and within the context of this model, 
we find that the contribution of the synchrotron self-Compton process
to the high-energy radiative output of FSRQs is negligible even in
their low-activity states.

\item We argue that the
difference between the low- and high-activity states in luminous blazars is due to
the different total kinetic power of the jet, most likely related to varying bulk Lorentz factor 
of the outflow within the blazar emission zone.

\end{enumerate}

\acknowledgments
The \textit{Fermi} LAT Collaboration acknowledges generous ongoing support
from a number of agencies and institutes that have supported both the
development and the operation of the LAT as well as scientific data analysis.
These include the National Aeronautics and Space Administration and the
Department of Energy in the United States, the Commissariat \`a l'Energie Atomique
and the Centre National de la Recherche Scientifique / Institut National de Physique
Nucl\'eaire et de Physique des Particules in France, the Agenzia Spaziale Italiana
and the Istituto Nazionale di Fisica Nucleare in Italy, the Ministry of Education,
Culture, Sports, Science and Technology (MEXT), High Energy Accelerator Research
Organization (KEK) and Japan Aerospace Exploration Agency (JAXA) in Japan, and
the K.~A.~Wallenberg Foundation, the Swedish Research Council and the
Swedish National Space Board in Sweden.

Additional support for science analysis during the operations phase is gratefully
acknowledged from the Istituto Nazionale di Astrofisica in Italy and the Centre National d'\'Etudes Spatiales in France.

\L S was partially supported by the Polish Ministry of Science and Higher Education through 
the project N N203 380336.


\end{document}